\documentclass[12pt]{article}
\usepackage[margin=1in]{geometry}
\usepackage[slantedGreek]{mathpazo}
\usepackage{graphicx}
\usepackage{amsfonts}
\usepackage{amssymb}
\usepackage{amsthm}
\usepackage{float}
\usepackage{amsmath}%
\usepackage{hyperref}
\usepackage{natbib}

\newtheorem{proposition}{Proposition}

\title{Some Bayesian Perspectives on Clinical Trials}
\author{
\makebox[.4\linewidth]{Alexandra Sokolova\thanks{Email: sokolova@ohsu.edu}}\\\textit{ Knight Cancer Institute }\\\textit{Hematology Oncology }\\\textit{ Oregon Health \& Science University}\\\and
\makebox[.4\linewidth]{Vadim Sokolov\thanks{Corresponding author. Email: vsokolov@gmu.edu}}\\\textit{ Department of Systems Engineering }\\\textit{  and Operations Research}\\\textit{ George Mason University}\\\and
\makebox[.4\linewidth]{Nick Polson\thanks{Email: nick.polson@chicagobooth.edu}}\\\textit{  Booth School of Business}\\\textit{  University of Chicago}
}
\date{First Draft: January 8, 2026 \\This Draft: \today}

\graphicspath{{./}}

\begin{document}
\maketitle
\begin{abstract}
We examine three landmark clinical trials---ECMO, CALGB~49907, and I-SPY~2---through a unified Bayesian framework connecting prior specification, sequential adaptation, and decision-theoretic optimisation. For ECMO, the posterior probability of treatment superiority is robust across the range of priors examined. For CALGB, predictive probability monitoring stopped enrolment at 633 instead of 1800 patients. For I-SPY~2, adaptive enrichment graduated nine of 23 arms to Phase~III.
These case studies motivate a methodological contribution: exact backward induction for two-arm binary trials, where Beta-Binomial conjugacy yields closed-form transitions on the integer lattice of success counts with no quadrature. A P\'olya-Gamma augmentation bridges this to covariate-adjusted logistic regression. Simulation reveals a fundamental tension: the optimal Bayesian design reduces expected sample sizes to 14--26 per arm (versus 42--100 for alternatives) but with substantially lower power. A calibrated variant embedding the declaration threshold in the terminal utility improves power while maintaining sample-size savings; varying the per-stage cost traces a power frontier for selecting the preferred operating point, with suitability highest in patient-sparing contexts such as rare diseases and paediatrics. The P\'olya-Gamma Laplace approximation is validated against exact calculations (mean absolute error below 0.01). We discuss implications for the 2026 FDA draft guidance on Bayesian methodology.
\end{abstract}


\section{Introduction}

The standard paradigm for clinical trial design---fix the sample size, randomise equally, analyse once at the end---was built to control the probability of approving an ineffective drug. But the framework prioritises error-rate control at the expense of within-trial patient outcomes: patients enrolled during the trial contribute to a statistical procedure whose conclusions are drawn only afterward. If early data show one treatment to be clearly inferior, patients continue receiving it. If the trial is underpowered, patients' participation generates inconclusive evidence.

The inefficiency is quantifiable. Consider a trial testing a new antihypertensive drug against standard care, with blood pressure reduction as the primary endpoint. The classical power calculation requires fixing the sample size to detect the minimum clinically important difference---say, 5~mmHg---with 90\% power at the two-sided 5\% level. With standard deviation $\sigma = 15$~mmHg, this yields approximately 190 patients per group, or 380 total. But suppose Phase~II data suggest the true effect is likely around 10~mmHg, not 5~mmHg, with uncertainty reflected in a prior distribution $\delta \sim N(10, 5^2)$. A Bayesian design that formally incorporates this prior information requires fewer patients to achieve comparable expected power (averaged over the prior on $\delta$), because the prior concentrates mass on plausible effect sizes rather than requiring protection against the worst case \citep{inoue2005}. The classical calculation ignores what is already known from Phase~II; the Bayesian calculation uses it.

The divergence reflects a deeper conceptual difference. The frequentist framework asks: ``If the null hypothesis were true and this experiment were repeated many times, would the decision rule produce false positives in fewer than 5\% of repetitions?'' The question is about error rates across hypothetical trials that will never be run. The Bayesian framework asks: ``Given the data actually observed and the prior information available, what do we believe about the treatment effect, and what decision maximises expected utility?'' When trials are monitored sequentially, this distinction becomes operational. The frequentist solution---alpha-spending functions that allocate type-I error across interim analyses---imposes a statistical penalty for looking at the data: to maintain overall $\alpha = 0.05$, early stopping boundaries require inflated critical values at initial looks (for example, $z \approx 4$ under O'Brien--Fleming spending), which can delay stopping even when evidence favours one arm. Futility boundaries partly mitigate this, but the constraint remains that each additional look consumes part of the error budget. The Bayesian posterior, by contrast, updates after every observation without a multiplicity penalty in the likelihood, because the stopping rule enters through the sampling model, not through the inference. This does not eliminate the need for calibration: Bayesian designs intended for confirmatory use still require pre-specified decision thresholds and simulation-based verification of operating characteristics (Section~\ref{sec:simulation}). Stopping rules based on posterior or predictive probabilities answer the question trialists actually face: should we stop now, or is the expected value of continuing positive?

The philosophical divide does not preclude practical agreement at the design stage---and the agreement is deeper than most practitioners realise. The frequentist formula for sample size determination is among the most familiar calculations in biostatistics. For a one-sided test of $H_0\colon \theta = \theta_0$ versus $H_1\colon \theta = \theta_1$ ($\theta_1 > \theta_0$) with Normal data and known variance $\sigma^2$:
\begin{equation}\label{eq:freq_ss}
n_F = (z_\alpha + z_\beta)^2 \left(\frac{\sigma}{\delta}\right)^2,
\end{equation}
where $\delta = \theta_1 - \theta_0$ is the target difference, $z_\alpha = \Phi^{-1}(1-\alpha)$, and $1 - \beta$ is the desired power. Every trialist knows this formula. What is less widely appreciated is that a Bayesian who frames the same question differently---seeking the smallest $n$ such that the expected utility of the experiment exceeds a threshold---arrives at the same sample size, not just for one $\delta$ but for all $\delta$ simultaneously.

\citet{inoue2005} formalised this correspondence by recasting sample-size determination as a decision problem with an explicit utility---a goal function---rather than conditional error rates. The Bayesian does not ask ``what are the type~I and type~II error rates?'' but ``what sample size minimises the expected cost of wrong decisions?'' Specifically, the Bayesian assigns prior probability $\pi$ to $H_0$ and $1 - \pi$ to $H_1$, and adopts a $0$-$1$-$K$ loss: no cost for a correct decision, cost $1$ for failing to reject $H_0$ when $H_1$ is true, and cost $K$ for rejecting $H_0$ when it is true. Minimising posterior expected loss yields a decision rule: do not reject $H_0$ whenever the posterior probability of $H_0$ exceeds $1/(1+K)$. For Normal data with known $\sigma^2$, this reduces to the cut-off
\[
\bar{x} \;\leq\; \frac{\sigma^2 \log\bigl(K\,\pi/(1-\pi)\bigr)}{n\,\delta} \;+\; \frac{\theta_1 + \theta_0}{2}\,.
\]
The goal function---equivalently, the expected utility of the experiment---is the utility-weighted probability of a correct decision (reducing to a pure classification rate when $K = 1$):
\begin{align}\label{eq:goal}
G_B(n) &= K\pi\,P_{\theta_0}(\text{accept}\;H_0) + (1-\pi)\,P_{\theta_1}(\text{reject}\;H_0) \nonumber\\
&= K\pi\,\Phi\!\left(\frac{\sigma\log\bigl(K\,\pi/(1\!-\!\pi)\bigr)}{\sqrt{n}\,\delta} + \frac{\delta\sqrt{n}}{2\sigma}\right) + (1\!-\!\pi)\,\Phi\!\left(\frac{\delta\sqrt{n}}{2\sigma} - \frac{\sigma\log\bigl(K\,\pi/(1\!-\!\pi)\bigr)}{\sqrt{n}\,\delta}\right),
\end{align}
where the second line follows by standardising $\bar{x}$ under each hypothesis. The Bayesian finds the smallest $n$ such that $G_B(n) \geq r^*$, a target level of expected utility. This is the Bayesian analogue of the frequentist power calculation---but framed entirely in terms of expected utility rather than conditional error rates. In this canonical setting, favourable type~I and type~II error rates emerge as a byproduct of maximising expected utility rather than as the primary design target.

Substituting $n = n_F$ from \eqref{eq:freq_ss} into \eqref{eq:goal}, every occurrence of $\delta$ and $\sigma$ cancels:
\begin{equation}\label{eq:inoue_const}
G_B(n_F) = K\pi\,\Phi\!\left(\frac{\log(K\pi/(1\!-\!\pi))}{|z_\alpha + z_\beta|} + \frac{|z_\alpha + z_\beta|}{2}\right) + (1\!-\!\pi)\,\Phi\!\left(\frac{|z_\alpha + z_\beta|}{2} - \frac{\log(K\pi/(1\!-\!\pi))}{|z_\alpha + z_\beta|}\right).
\end{equation}
The right-hand side depends only on $(\alpha, \beta, \pi, K)$, not on $\delta$ or $\sigma$. With $\pi = 1/2$ and $K = 1$ (symmetric prior and loss), the expression simplifies to $\Phi\bigl((z_\alpha + z_\beta)/2\bigr)$, which at $\alpha = 0.05$ and $\beta = 0.10$ gives $r^* = 0.928$. A frequentist who calculates $n_F = 857$ for $\sigma = 1$, $\delta = 0.10$ and a Bayesian who requires $r^* = 0.928$ will agree on $n = 857$---and they will continue to agree at $\delta = 0.05$ ($n = 3{,}426$), $\delta = 0.20$ ($n = 214$), and at every other $\delta$. The choice of $\delta$ in the frequentist calculation implicitly encodes the same information as the Bayesian prior: both determine the signal-to-noise ratio at which the study becomes conclusive.

The result extends to composite hypotheses $H_0\colon \theta \leq \theta_0$ versus $H_1\colon \theta > \theta_0$ under a Normal prior $\theta \sim N(\theta_0, \tau^2)$, provided $\tau^2 = C\delta^2$ for a constant $C > 0$---that is, the prior standard deviation scales with the target effect. It also extends to Lindley information as the Bayesian goal, where the information target $\mathcal{I}^* = \tfrac{1}{2}\log(1 + C(z_\alpha + z_\beta)^2)$ is again constant in $\delta$, and to Bernoulli data, where the correspondence holds approximately.

The practical implication is that, in these canonical sample-size settings, Bayesian and frequentist approaches are not in conflict---they are different parametrisations of the same calculation. Every trialist knows the power formula \eqref{eq:freq_ss}, but few recognise that the familiar inputs $(\alpha, \beta, \delta)$ implicitly encode a prior and a utility: the ``relevant difference'' $\delta$ plays the role of the prior location, and the error rates $\alpha$ and $\beta$ map to the asymmetric loss $K$ and classification threshold $r^*$. The power calculation is not wrong---it is a goal function in disguise.

The tension we develop in this paper is therefore not between Bayesian and frequentist philosophies per se, but between designs that optimise expected utility and designs that optimise frequentist operating characteristics. At the design stage, as \eqref{eq:inoue_const} shows for the canonical Normal setting with known variance, the two coincide. It is in sequential monitoring that they diverge. The frequentist who monitors a trial sequentially cannot simply look at the accumulating data: each interim analysis consumes part of the type~I error budget, the alpha-spending function must be pre-specified, and modifying it mid-trial requires a protocol amendment. In practice, this means that even when early data strongly favour one arm, the frequentist framework imposes a statistical cost for having looked. The Bayesian posterior, by contrast, updates after every observation without such a penalty. In the sequential setting, the organising principle remains expected utility---not type~I and type~II error---and the goal function \eqref{eq:goal} generalises naturally by conditioning on the data observed so far. This asymmetry---the freedom to inspect data at any time, optimising expected utility at each step---is the operational advantage that motivates the sequential designs we develop next.

The foundational idea is simple: probability is a measure of uncertainty, updated coherently as data arrive. This principle has four consequences for clinical trials that we develop in this paper:
\begin{enumerate}
\item[(a)] \textbf{Priors encode what is already known.} Treatment effects in clinical trials are small---typically a few percentage points of absolute risk reduction. This prior knowledge, ignored by noninformative analyses, can improve both efficiency and calibration when formalised correctly. More broadly, the posterior precision at any sample size equals the sum of prior precision and data precision. With an informative prior, the posterior starts concentrated and tightens with each observation; without one, the first observations are wasted re-learning what was already known.
\item[(b)] \textbf{Small samples suffice for strong conclusions.} Because the posterior accumulates evidence continuously, a trial can stop as soon as the posterior probability of superiority (or futility) crosses a decision threshold---whether that takes 12 patients or 600. The Bayesian framework imposes no likelihood-based penalty for early inspection---the posterior is valid regardless of the stopping rule---so evidence that arrives quickly is acted on quickly. (The frequentist operating characteristics of repeated monitoring still require calibration; see Section~\ref{sec:simulation}.) With an informative prior, fewer observations are needed to cross any given threshold, because the prior has already contributed information equivalent to its effective sample size.
\item[(c)] \textbf{Sequential decisions can be made optimally.} When utility depends on sufficient statistics, the stopping problem reduces to a computable table via backward induction. Combined with Thompson's (1933) probability-matching rule, this yields adaptive designs that balance learning against treating.
\item[(d)] \textbf{Trial design is a decision problem.} The choice of dose, the rule for stopping, and the allocation of patients are decisions under uncertainty. A utility function that jointly values efficacy, toxicity, and information gain provides a coherent framework for making them.
\end{enumerate}
The practical upshot is that Bayesian trials can reach definitive conclusions---in either direction---with fewer patients than fixed-sample designs, though the magnitude of the saving depends on the utility function and comes with trade-offs in power and type~I error that we quantify in Section~\ref{sec:simulation}. Informative priors amplify the sample-size advantage. The case studies in this paper illustrate the point repeatedly: ECMO reached a strong conclusion with 12 patients (Section~\ref{sec:cases}), CALGB~49907 stopped 65\% early, and a hierarchical model applied to multi-centre data produces a 98.5\% posterior probability of treatment efficacy from centres with as few as five patients per arm (Section~\ref{sec:priors}).

We are not the first to advocate Bayesian clinical trials. \citet{novick1965} introduced Bayesian analysis of clinical data using conjugate priors for categorical outcomes. \citet{berry2010} provide a comprehensive treatment of Bayesian adaptive methods. \citet{berry2025} reviews the state of the art. Our contributions are threefold. First, we develop exact backward induction for two-arm binary trials, exploiting the Beta-Binomial conjugacy to obtain closed-form transition probabilities on the integer lattice of success counts. Unlike the Normal model of \citet{brockwell2003}, which requires numerical quadrature on a continuous grid, the binary model yields exact optimal stopping rules with no discretisation error. A calibrated variant embeds the declaration threshold directly in the terminal utility, yielding a power frontier that makes the sample-size--power trade-off explicit and tunable. Second, we show how P\'olya-Gamma augmentation \citep{polson2013} bridges this framework to logistic regression with covariate adjustment, and validate the resulting Laplace approximation against exact Beta posterior calculations. Third, we demonstrate through three landmark trials how the Bayesian principles of prior specification, sequential adaptation, and decision-theoretic optimisation interact in practice: informative priors sharpen inference when data are sparse (ECMO), predictive probabilities enable early stopping that saves hundreds of patients (CALGB~49907), and utility-based adaptive enrichment accelerates drug development in heterogeneous populations (I-SPY~2). For the ECMO trial, we provide a dedicated backward induction re-analysis with asymmetric historical priors.

Section~\ref{sec:priors} discusses prior specification. Section~\ref{sec:sequential} covers sequential adaptive designs, including exact binary backward induction, the P\'olya-Gamma bridge, and a simulation study with power frontier analysis. Section~\ref{sec:decision} develops the decision-theoretic framework. Section~\ref{sec:cases} presents case studies, including a backward induction re-analysis of the ECMO trial. Section~\ref{sec:regulatory} discusses the regulatory landscape. Section~\ref{sec:discussion} concludes.

\section{The Role of Priors}\label{sec:priors}

The prior distribution is the defining feature of Bayesian inference and the source of most controversy surrounding it. In clinical trials, the controversy is misplaced: we know far more about treatment effects before a trial begins than a noninformative prior would suggest.

\subsection{Why default priors fail for treatment effects}

Consider a two-arm trial comparing a new treatment to control, with binary outcome (response/no response). Let $\theta_1$ and $\theta_0$ denote the response probabilities under treatment and control, respectively. The treatment effect is $\delta = \theta_1 - \theta_0$.

Jeffreys' prior for the binomial proportion is $\text{Beta}(1/2, 1/2)$, which places substantial mass near 0 and 1. Under independent Jeffreys priors for $\theta_1$ and $\theta_0$, the implied prior on $\delta$ is U-shaped, concentrating mass near $\pm 1$. This contradicts everything known about drug development: treatment effects are small. Among drugs that reach Phase III, absolute risk reductions are typically modest---often on the order of a few percentage points \citep{berry2012}. A prior that places most mass on treatment effects near $\pm 100\%$ is not noninformative---it is misinformative.

The uniform prior $\text{Beta}(1,1)$ is less extreme but still problematic. It implies that a response rate of 95\% is as likely as 50\% a priori, and that the treatment effect $\delta$ has prior standard deviation of approximately 0.41---far larger than any plausible effect in most therapeutic areas. From a regularisation perspective, the choice of prior is a modelling decision with direct inferential consequences: a uniform prior corresponds to maximum likelihood, while informative priors provide the Bayesian analogue of penalised estimation \citep{polsonsokolov2019}.

\begin{figure}[t]
\centering
\includegraphics[width=0.85\textwidth]{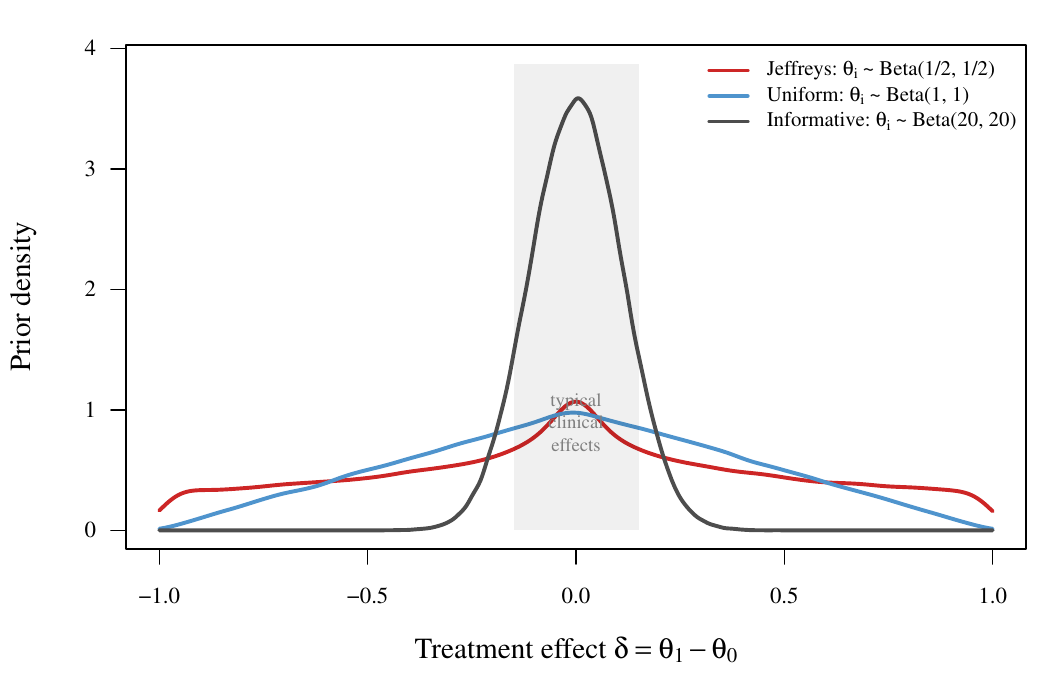}
\caption{Implied prior density on the treatment effect $\delta = \theta_1 - \theta_0$ under three choices of independent priors for the success probabilities $\theta_0$ and $\theta_1$. The shaded band marks the range of typical clinical effects ($|\delta| \leq 0.15$). Jeffreys priors (red) concentrate mass near $\pm 1$; uniform priors (blue) spread mass broadly; an informative prior (grey) concentrates mass where treatment effects are known to lie.}
\label{fig:priors}
\end{figure}

Figure~\ref{fig:priors} makes the problem visible: the implied prior on $\delta$ under Jeffreys priors is U-shaped, placing most mass on extreme effects near $\pm 1$; under uniform priors it is triangular, still far too diffuse; only an informative prior concentrates mass where clinical effects actually live.

\citet{novick1965} recognised this issue and proposed ``logical probability'' priors calibrated to prior data. Their natural conjugate Bayes densities express prior information as equivalent to $(m, z)$ prior observations with $z$ successes, so the prior $\text{Beta}(z+1, m-z+1)$ has the interpretation of $m$ prior patients. This parametrisation makes the prior's influence transparent: it adds $m$ pseudo-observations to the data. The effective sample size concept was later formalised by \citet{morita2008}.

\subsection{Informative priors from historical data}

A more principled approach uses historical data to construct informative priors. \citet{racinepoon1986} demonstrated this systematically across four areas of pharmaceutical research at Ciba-Geigy: LD50 (median lethal dose) experiments, crossover trials, bioequivalence assessment, and population pharmacokinetics. For LD50 experiments, where small sample sizes cause fiducial intervals to degenerate to the entire real line, Bayesian priors calibrated to toxicity class boundaries produced posterior densities that were both well-defined and practically useful. For population pharmacokinetic models---hierarchical nonlinear random effects models where individual parameters $\theta_i$ vary across patients---the Bayesian framework borrows strength across individuals, enabling prediction of a new patient's drug concentration profile from as few as two blood draws when combined with population data. If a drug's mechanism of action is understood, prior knowledge about the control arm response rate is often substantial. For dose-response relationships, the LD50 (the dose producing 50\% response) provides a natural anchor: a prior on the dose-response curve centred at the LD50 with variance reflecting uncertainty about the curve's slope encodes genuine pharmacological knowledge.

A fundamental difficulty in specifying priors for dose-response models is that regression coefficients are abstract and model-dependent: the meaning of $\beta$ in a logistic model differs from its meaning in a probit model, so eliciting priors directly on $\beta$ is unnatural. \citet{bedrick1996} resolved this by introducing \emph{conditional means priors}: the investigator specifies prior distributions on observable quantities---the probability of response at selected covariate values---and the prior on $\beta$ is induced through the link function. For a binomial regression with $p$ predictors, one specifies independent $\text{Beta}(a_{1i}, a_{2i})$ priors on the success probabilities $\tilde{m}_i$ at $p$ chosen covariate vectors $\tilde{\mathbf{x}}_i$, and the induced prior on $\beta$ takes the form
\begin{equation}\label{eq:cmp}
\pi(\beta) \propto \prod_{i=1}^p F(\tilde{\mathbf{x}}_i'\beta)^{a_{1i}-1}[1 - F(\tilde{\mathbf{x}}_i'\beta)]^{a_{2i}-1} f(\tilde{\mathbf{x}}_i'\beta),
\end{equation}
where $F$ is the CDF associated with the link function and $f$ its density. In the Challenger O-ring analysis, \citet{bedrick1996} modelled the probability that any O-ring fails as a function of temperature via $g(m_i) = \beta_0 + \beta_1(\tau_i - \bar{\tau})$ and specified independent beta priors on the failure probabilities at two temperatures: $\tilde{m}_1 \sim \text{Beta}(1, 0.577)$ at $55^\circ$F (giving $\Pr[\tilde{m}_1 > 1/2] = 2/3$, reflecting a belief that failure is likely at low temperatures) and $\tilde{m}_2 \sim \text{Beta}(0.577, 1)$ at $75^\circ$F (giving $\Pr[\tilde{m}_2 < 1/2] = 2/3$). These priors on observable failure rates are easy to elicit; the induced prior on $(\beta_0, \beta_1)$ via \eqref{eq:cmp} then depends on the link function, so the same scientific beliefs produce different coefficient priors for logistic versus probit models without requiring the investigator to think about model-dependent abstractions. \citet{racinepoon1986} used a similar idea informally, specifying independent beta priors on two success probabilities to induce bivariate priors on probit regression coefficients in bioequivalence studies at Ciba-Geigy; \citet{bedrick1996} formalised and generalised the approach.

\begin{figure}[t]
\centering
\includegraphics[width=0.95\textwidth]{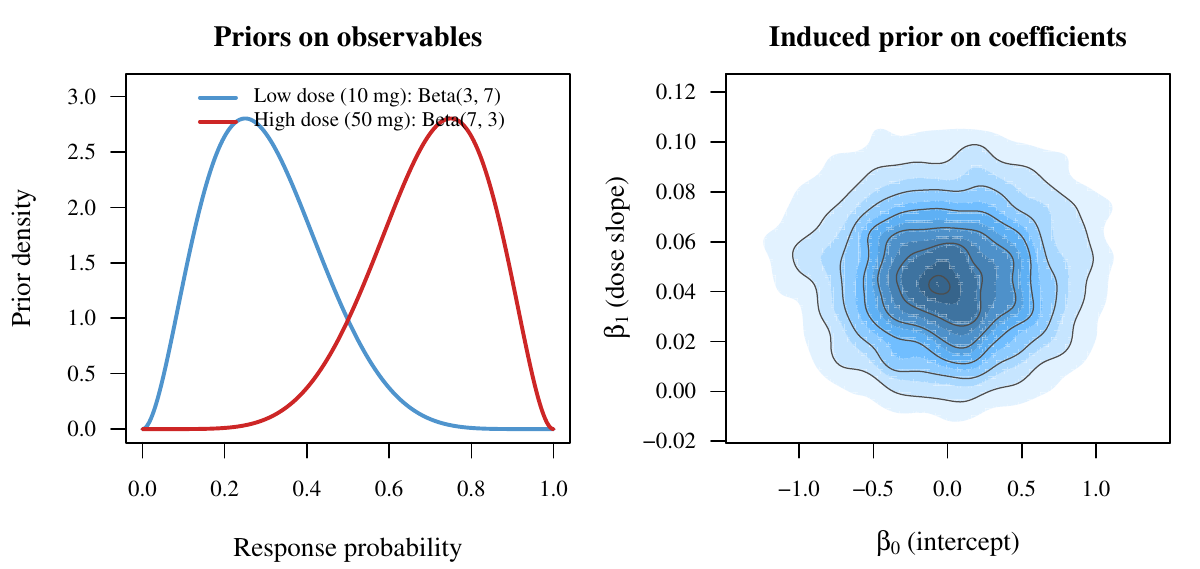}
\caption{Conditional means prior for a dose-response model. \emph{Left:} the clinician specifies Beta priors on the response probability at a low dose (10~mg, blue) and a high dose (50~mg, red). \emph{Right:} the induced joint prior on the logistic regression coefficients $(\beta_0, \beta_1)$ via \eqref{eq:cmp}. The prior on coefficients is proper and concentrated, even though the clinician never reasoned about regression parameters directly.}
\label{fig:cmp}
\end{figure}

For clinical trials, the translation is direct: a clinician can state beliefs about response rates at clinically meaningful doses (``at 10~mg, I expect 30--50\% response; at 50~mg, perhaps 60--80\%'') and the conditional means prior machinery converts these into a coherent prior on regression coefficients for whatever dose-response model is chosen. Figure~\ref{fig:cmp} illustrates this for a dose-response trial: independent Beta priors on response probabilities at two doses (left panel) induce through the logistic link a proper, concentrated joint prior on the intercept and slope (right panel).

For logistic regression specifically, the induced prior is equivalent to a data augmentation prior---pseudo-observations at the design points with fractional sample sizes---connecting directly to the P\'olya-Gamma scheme below. \citet{hanson2014} extended this through informative $g$-priors for logistic regression, where the investigator specifies only a prior on the overall population success probability and the $g$-prior structure induces a proper joint prior on all coefficients. Together with \citet{chaloner1993}, who characterised optimal Bayesian designs for dose-response experiments under such priors, these methods enable the trialist to specify beliefs about response rates at clinically meaningful doses, let the link function induce priors on model parameters, and design the experiment to maximise information about the LD50 or other quantities of interest.

The FDA's 2026 draft guidance on Bayesian methodology \citep{fda2026} identifies several settings where informative priors from historical data have supported regulatory submissions: borrowing from previous clinical trials of the same drug, augmenting concurrent controls with external data, paediatric extrapolation from adult trials, and borrowing across similar disease subtypes. In each case, the prior's influence must be quantified---for example, through its effective sample size---and sensitivity analyses must demonstrate robustness to prior-data conflict.

Dynamic borrowing methods, where the degree of information transfer depends on observed consistency between historical and current data, mitigate the risk of static priors that do not respond to conflict. The Bayesian hierarchical framework provides a natural mechanism: the between-study variance parameter $\tau^2$ governs borrowing, with large $\tau^2$ producing sceptical priors that are quickly overwhelmed by current data.

\subsection{P\'olya-Gamma augmentation}

Many clinical trial endpoints are binary or categorical, leading to logistic regression models for treatment effects adjusted for covariates. Bayesian inference for logistic regression historically required Metropolis-Hastings sampling, which is slow and requires careful tuning.

\citet{polson2013} introduced the P\'olya-Gamma data augmentation scheme, a special case of the variance-mean mixture framework for non-Gaussian regression \citep{polsonscott2013da}, which renders the posterior for logistic regression conditionally Gaussian. The key identity is:
\begin{equation}\label{eq:pg}
\frac{(e^\psi)^a}{(1+e^\psi)^b} = 2^{-b} e^{\kappa \psi} \int_0^\infty e^{-\omega \psi^2/2} p(\omega) \, d\omega,
\end{equation}
where $\kappa = a - b/2$ and $\omega \sim \text{PG}(b, 0)$ is a P\'olya-Gamma random variable. Conditional on $\omega$, the log-odds $\psi$ has a Gaussian full conditional, enabling a simple Gibbs sampler that alternates between drawing $\omega$ from a P\'olya-Gamma distribution and drawing regression coefficients from a multivariate Gaussian.

For clinical trials with binary endpoints and moderate covariate adjustment, this scheme produces posterior samples via a Gibbs sampler with no tuning parameters, making real-time Bayesian monitoring computationally~straightforward. The approach extends naturally to multinomial and negative binomial models.

\paragraph{Multi-centre trials.} A direct application arises in multi-centre binary-response studies, where the same treatment comparison is conducted across $N$ centres with potentially different response rates. Let $y_{ij}$ denote the number of successes among $n_{ij}$ patients at centre $i$ on arm $j \in \{1,2\}$ (treatment, control), with log-odds $\psi_{ij} = \log\{p_{ij}/(1-p_{ij})\}$. The hierarchical logistic-normal model places a bivariate normal prior on each centre's log-odds vector:
\begin{equation}\label{eq:multicentre}
\psi_i = (\psi_{i1}, \psi_{i2})' \mid \mu, \Sigma \sim N(\mu, \Sigma), \qquad p(\mu) \propto 1, \qquad \Sigma \sim \text{IW}(d, B),
\end{equation}
where $\mu = (\mu_1, \mu_2)'$ represents the overall treatment and control log-odds, the improper flat prior on $\mu$ is standard, and $\Sigma$ captures between-centre variability. \citet{skene1990} proposed this model and analysed it via numerical integration; \citet{polsonscott2011tables} showed that P\'olya-Gamma augmentation renders the posterior conditionally conjugate, yielding a tuning-free Gibbs sampler.

Applying the P\'olya-Gamma identity \eqref{eq:pg} to each binomial likelihood factor $p(y_{ij} \mid \psi_{ij}) \propto e^{y_{ij}\psi_{ij}}/(1 + e^{\psi_{ij}})^{n_{ij}}$ introduces latent variables $\omega_{ij} \mid \psi_{ij} \sim \text{PG}(n_{ij}, \psi_{ij})$. Conditional on $\boldsymbol{\omega}$, the centre-level log-odds have a Gaussian full conditional:
\begin{equation}\label{eq:multicentre_psi}
\psi_i \mid \boldsymbol{\omega}_i, \mu, \Sigma, D \sim N(\mathbf{m}_i, \mathbf{V}_i), \qquad \mathbf{V}_i^{-1} = \boldsymbol{\Omega}_i + \Sigma^{-1}, \quad \mathbf{m}_i = \mathbf{V}_i(\boldsymbol{\kappa}_i + \Sigma^{-1}\mu),
\end{equation}
where $\boldsymbol{\Omega}_i = \text{diag}(\omega_{i1}, \omega_{i2})$ and $\boldsymbol{\kappa}_i = (y_{i1} - n_{i1}/2, \; y_{i2} - n_{i2}/2)'$. The hyperparameters update conjugately: $\mu \mid \Psi, \Sigma \sim N(\bar{\psi}, \Sigma/N)$ and $\Sigma \mid \Psi, \mu \sim \text{IW}(d + N, B + \sum_{i=1}^N (\psi_i - \mu)(\psi_i - \mu)')$.

The structure of \eqref{eq:multicentre_psi} also clarifies the connection to the conditional means prior \eqref{eq:cmp}. If one replaces the logistic-normal prior in \eqref{eq:multicentre} by independent $\text{Beta}(a_{ij}, b_{ij})$ priors on the success probabilities $p_{ij}$---equivalently, independent Fisher $Z(a_{ij}, b_{ij})$ priors on the log-odds $\psi_{ij}$---then P\'olya-Gamma augmentation absorbs these priors into the Gaussian conditional as $a_{ij} + b_{ij}$ pseudo-observations \citep{polsonscott2011tables}. The elicited $\text{Beta}$ hyperparameters at design points become pseudo-data that enter the Gibbs sampler on the same footing as observed counts, exactly as in the Bedrick--Christensen--Johnson framework. This provides a computationally trivial route from the conditional means prior specification of Section~\ref{sec:priors} to full posterior inference, whether the hierarchical coupling across centres is through a shared logistic-normal prior on $\psi$ or through independent Z-priors with common hyperparameters.

\begin{figure}[H]
\centering
\includegraphics[width=0.75\textwidth]{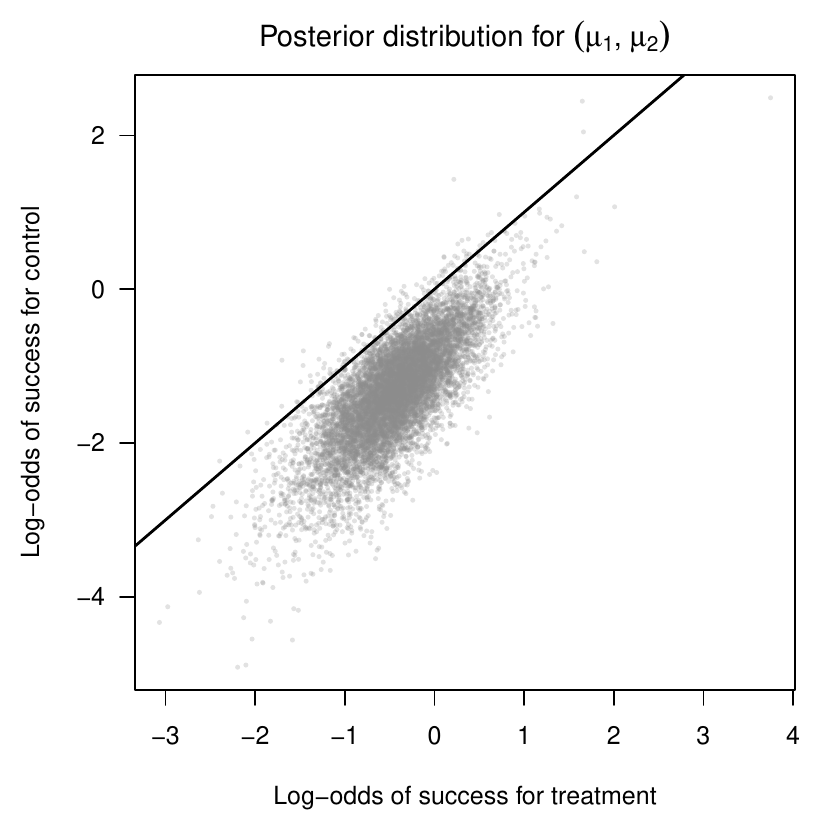}
\caption{Joint posterior for $\mu = (\mu_1, \mu_2)'$ in the hierarchical logistic-normal model \eqref{eq:multicentre} fitted to the eight-centre topical cream data of \citet{skene1990} via P\'olya-Gamma Gibbs sampling. Each grey dot is a posterior draw; the diagonal line marks $\mu_1 = \mu_2$ (equal treatment and control). The mass below the diagonal ($\Pr(\mu_1 > \mu_2 \mid \text{data}) = 0.985$) indicates strong evidence of treatment efficacy despite small per-centre samples.}
\label{fig:mupost}
\end{figure}

Figure~\ref{fig:mupost} illustrates the general small-sample advantage on data from an eight-centre study of topical cream effectiveness \citep{skene1990}. Individual centres are small---5 to 37 patients per arm---and most are individually inconclusive (although one or two reach nominal significance by Fisher's exact test, none survive a Bonferroni correction for eight comparisons). Yet the hierarchical model borrows strength across centres, and the posterior for $\mu$ places 98.5\% of its mass in the region $\mu_1 > \mu_2$ (below the diagonal in the figure), reaching a strong pooled conclusion that no single centre could support on its own. The Bayesian shrinkage is especially important for centres~5 and~6, where zero control-arm successes would produce infinite maximum-likelihood log-odds estimates; the hierarchical prior regularises these towards the population mean. The mechanism is visible in the precision matrix of \eqref{eq:multicentre_psi}: each centre's posterior precision $\mathbf{V}_i^{-1} = \boldsymbol{\Omega}_i + \Sigma^{-1}$ combines local data (through $\boldsymbol{\Omega}_i$, which grows with sample size) with information from all other centres (through $\Sigma^{-1}$, which acts as an informative prior). A centre with few patients has small $\boldsymbol{\Omega}_i$ and is therefore pulled strongly towards the population mean $\mu$; a centre with many patients retains its own estimate. This is why the Bayesian framework reaches conclusions quickly in either direction: centres with clear evidence for or against the treatment contribute that evidence to the population-level inference, while centres with ambiguous data borrow from the collective. With an informative prior on $\mu$ calibrated to historical response rates---as in the Bedrick framework of Section~\ref{sec:priors}---the borrowing starts before the first patient is enrolled, enabling each centre to reach a decision with even fewer observations.

\section{Sequential Adaptive Designs}\label{sec:sequential}

The ethical case for adaptive designs is simple: if accumulating data show one treatment to be inferior, continuing to assign patients to it is unjustifiable. The statistical case is that sequential allocation can improve efficiency, reducing the number of patients needed to reach a conclusion.

The frequentist approach to interim monitoring is group sequential testing. \citet{pocock1977} proposed equal critical values at each interim look; \citet{obrienfleming1979} proposed conservative early boundaries that spend little alpha initially. \citet{landemets1983} unified these through the alpha-spending function $\alpha^*(t)$, which allocates the overall type-I error rate across information fractions $t \in [0,1]$. The spending function need not be specified until the interim analysis is conducted, giving the trialist flexibility in the timing and number of looks.

The Bayesian sequential framework differs in three respects. First, it conditions on the data actually observed rather than on a pre-specified schedule of looks, so inference under the assumed model is valid regardless of when or why the data were examined. Second, the stopping rule is based on predictive or posterior probabilities that answer the clinical question directly (``what is the probability that this treatment is superior?'') rather than on $p$-values adjusted for multiplicity. Third, the Bayesian framework extends naturally to response-adaptive randomisation and multi-arm designs; frequentist group sequential methods can accommodate multiple arms and adaptive allocation, but doing so requires increasingly elaborate multiplicity adjustments, whereas the Bayesian approach handles these extensions through the posterior, though confirmatory trials still require simulation-based calibration of operating characteristics including type~I error.

\subsection{Thompson sampling}

\citet{thompson1933} proposed the earliest response-adaptive randomisation rule. Given two treatments with unknown success probabilities $\tilde{p}_1$ and $\tilde{p}_2$, and prior data consisting of $r_i$ successes in $n_i$ patients for treatment $i$, Thompson computed
\begin{equation}\label{eq:thompson}
P(\tilde{p}_2 > \tilde{p}_1 \mid \text{data}) = \frac{\sum_{\alpha=0}^{r_2} \binom{r_1+r_2-\alpha}{r_1} \binom{s_1+s_2+1+\alpha}{s_1}}{\binom{n_1+n_2+2}{n_1+1}},
\end{equation}
where $s_i = n_i - r_i$, and proposed assigning the next patient to treatment 2 with this probability. This ``probability matching'' rule assigns patients to the treatment that is more likely to be better, in proportion to the evidence.

Thompson's motivation was explicitly ethical: ``a considerable saving of individuals otherwise sacrificed to the inferior treatment might be effected.'' Under uniform priors, the posterior for each treatment's success rate is $\text{Beta}(r_i+1, s_i+1)$, and the allocation probabilities in \eqref{eq:thompson} follow from the integral $\int_0^1 \int_0^{\theta_2} \pi(\theta_2|\text{data}) \pi(\theta_1|\text{data}) \, d\theta_1 \, d\theta_2$.

Thompson sampling was rediscovered by the technology industry in the 2000s and is now widely adopted for online optimisation problems \citep{scott2010}. Advances in Bayesian computation made randomised probability matching easy to implement for virtually any payoff distribution. The adoption of Thompson sampling for allocating web traffic---where the stakes are advertising revenue---suggests that the case for adaptive allocation is at least as strong in clinical trials, where the stakes are patient outcomes.

\begin{figure}[t]
\centering
\includegraphics[width=0.85\textwidth]{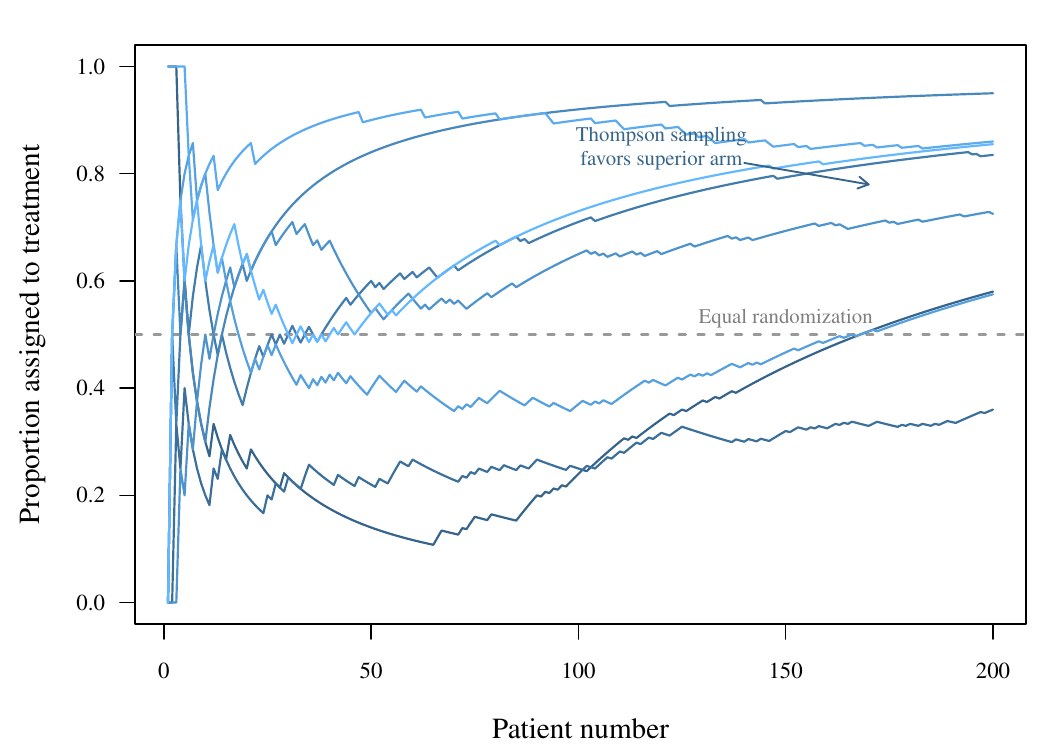}
\caption{Thompson sampling in a two-arm trial ($p_0 = 0.30$, $p_1 = 0.45$). Each curve shows the running proportion of patients assigned to the treatment arm across eight independent simulations. Under equal randomisation (dashed line), half of all patients receive the inferior control. Thompson sampling progressively shifts allocation toward the superior arm as evidence accumulates.}
\label{fig:thompson}
\end{figure}

Figure~\ref{fig:thompson} shows simulated Thompson sampling paths for a two-arm trial with control success rate $p_0 = 0.30$ and treatment success rate $p_1 = 0.45$. The allocation proportion drifts above the equal-randomisation line as evidence accumulates, assigning more patients to the superior arm.

The \citet{thallwathen2007} generalisation raises the posterior probabilities to a power $\tau$ before normalising:
\begin{equation}\label{eq:thallwathen}
r_{k,n} = \frac{[p_{k,n}]^\tau}{\sum_{j=1}^K [p_{j,n}]^\tau},
\end{equation}
where $p_{k,n} = \Pr(\pi_k > \pi_j \text{ for all } j \neq k \mid D_n)$ is the posterior probability that arm $k$ is best after $n$ patients, and $K$ is the number of arms. When $\tau = 0$, this reduces to equal randomisation; when $\tau \to \infty$, it assigns all patients to the current leader. The tuning parameter $\tau$ controls the exploration-exploitation trade-off inherent in the multi-armed bandit formulation of clinical trials.

\subsection{Predictive probabilities and stopping rules}

A trial's most consequential decision is when to stop. Bayesian predictive probabilities answer the question trial teams actually face: given current data, what is the probability that the trial will reach a definitive conclusion if continued? The idea has an early precedent in bioequivalence testing: \citet{racinepoon1987} proposed a two-stage procedure in which first-stage data are used to compute the predictive probability of establishing bioequivalence in a second stage, and the second-stage sample size $n_2$ is chosen---or the study abandoned---accordingly. Their simulations showed that the two-stage design roughly doubles the success rate of single-stage studies at a comparable average sample size.

Let $Z_n$ denote the test statistic at the interim analysis with information fraction $r = I_n / I_N$, where $I_n$ is the current information and $I_N$ is the information at the planned final analysis. \citet{marion2025} show that the predictive probability of trial success can be approximated as
\begin{equation}\label{eq:pp}
\text{PP}(p_n, r, \alpha) = \Phi\left(\frac{\Phi^{-1}(1-p_n) - \Phi^{-1}(1-\alpha)\sqrt{r}}{\sqrt{1-r}}\right),
\end{equation}
where $p_n$ is the one-sided $p$-value at the interim (equivalently, $p_n = 1 - \mathcal{P}_n$ when $\mathcal{P}_n$ is the Bayesian posterior probability of superiority under a flat prior), $r = I_n / I_N$ is the observed information fraction, and $\alpha$ is the one-sided significance threshold for the final analysis. The approximation is exact when the test statistic follows a Gaussian information-monitoring process with known $r$ and a flat prior, and accurate across a range of endpoint types---binary, time-to-event, and ordinal---as verified by simulation against full Monte Carlo imputation \citep{marion2025}.

The computational advantage is substantial: where full predictive probability calculations require nested simulations (simulating future data within each simulated interim analysis), the approximation in \eqref{eq:pp} requires only the current $p$-value and the information fraction. This enables trial simulation with orders-of-magnitude speedup.

\subsection{Backward induction on sufficient statistics}

The optimal sequential stopping rule can in principle be found by backward induction: starting at the maximum horizon, compute the optimal decision for every possible data configuration, then work backward. For general data, the decision tree grows exponentially and the computation is intractable.

\citet{christen2003}, building on \citet{berry2000} and \citet{carlin1998}, identified a key structural insight: when the utility function depends on data only through a sufficient statistic, the exponentially growing decision tree collapses to a polynomial-size table.

Consider a sequential experiment with horizon $T$, data $\mathbf{X}_t = (X_1, \ldots, X_{n+t})$ after $t$ additional observations, and utility for stopping at time $t$:
\begin{equation}\label{eq:utility}
u(1, \mathbf{X}_t) = a \, S(t) + b(T - t),
\end{equation}
where $S(t) = \sum_{l=1}^{n+t} X_l$ is the running total and $a, b$ are reward and cost parameters. The utility of continuing is the expected optimal utility at the next step:
\begin{equation}\label{eq:continue}
u(0, \mathbf{x}_t) = \sum_{\mathbf{x}_{t+1}} U(\mathbf{x}_{t+1}) \, P(\mathbf{X}_{t+1} = \mathbf{x}_{t+1} \mid \mathbf{X}_t = \mathbf{x}_t),
\end{equation}
where $U(\mathbf{x}_{t+1}) = \max\{u(1, \mathbf{x}_{t+1}), u(0, \mathbf{x}_{t+1})\}$. Because $(S(t), V(t))$ with $V(t) = (n+t)S(t) - \sum_{l=1}^{n+t}(l-1)X_l$ is a sufficient statistic for the model, and the utility depends on $\mathbf{X}_t$ only through $S(t)$, the quantity $u(0, \mathbf{x}_t)$ depends on $\mathbf{x}_t$ only through $(S(t), V(t))$. The decision tree reduces from exponential in $T$ to a three-dimensional table indexed by $(t, s, v)$, with backward recursion:
\begin{multline}\label{eq:backward}
u_t(0, s, v) = \sum_{h=0}^{\infty} U_{t+1}(s+h, v+s+h) \\
\times P(S(t+1)=s+h, V(t+1)=v+s+h \mid S(t)=s, V(t)=v).
\end{multline}
The transition probabilities are estimated by Monte Carlo simulation from the posterior predictive distribution. In practice, the table can be further reduced to two dimensions $(t, s)$ when the accumulation curve is near its plateau and $V(t)$ is approximately determined by $S(t)$.

\citet{carlin1998} applied this sufficient-statistic reduction to clinical trials directly, developing sequential decision-theoretic methods for treatment allocation. Their approach demonstrated that backward induction---the ``theoretically sound and generally accepted procedure for Bayesian sequential analysis'' that is ``hardly ever implemented due to its great complexities'' \citep{berger1985}---becomes feasible when the problem structure is exploited.

\subsection{The Normal model and gridding: a worked example}

We develop the backward induction in full for the clinical trial problem of \citet{brockwell2003}, following \citet{carlin1998}. This example makes the sufficient-statistic reduction concrete and illustrates the gridding algorithm that makes optimal sequential design computationally tractable.

\paragraph{Setup.} Patients arrive sequentially, and the outcome $Z_k$ for patient $k$ given treatment effect $\theta$ satisfies
\begin{equation}\label{eq:normal_model}
Z_k \mid \theta \sim N(\theta, \sigma^2), \qquad \theta \sim N(0, \sigma_0^2),
\end{equation}
with $\sigma^2$ known. The decision at each stage is either to stop and choose a terminal action $d \in \{d_0, d_1\}$ (declare for control or treatment) or to continue sampling at cost $c$ per patient. The loss for terminal decision $d_1$ (favour treatment) when the true effect is $\theta$ is $R(\theta, d_1)$, and similarly for $d_0$; a natural choice is $R(\theta, d_1) = -\theta$ (reward equals effect size if treatment is chosen) and $R(\theta, d_0) = 0$.

\paragraph{Sufficient statistics.} After $n$ observations, the posterior is $\theta \mid z_1, \ldots, z_n \sim N(S_n, \psi_n^2)$, where
\begin{equation}\label{eq:normal_ss}
S_n = E[\theta \mid z_1, \ldots, z_n] = \psi_n^2 \left(\frac{S_{n-1}}{\psi_{n-1}^2} + \frac{z_n}{\sigma^2}\right), \qquad \psi_n^2 = \left(\frac{1}{\psi_{n-1}^2} + \frac{1}{\sigma^2}\right)^{-1} = \frac{\sigma^2 \psi_{n-1}^2}{\sigma^2 + \psi_{n-1}^2}.
\end{equation}
Since $\psi_n^2$ is deterministic (it depends only on $n$, $\sigma^2$, and $\sigma_0^2$), the state of the trial after $n$ patients is fully described by the scalar $S_n$. Decision rules, expected losses, and continuation values all depend on data only through $S_n$.

\paragraph{Predictive transitions.} The next observation has predictive distribution
\begin{equation}\label{eq:predictive}
Z_{n+1} \mid S_n \sim N(S_n, \, \sigma^2 + \psi_n^2),
\end{equation}
and since $S_{n+1}$ is a linear function of $Z_{n+1}$, the updated posterior mean is also Gaussian:
\begin{equation}\label{eq:transition}
S_{n+1} \mid S_n \sim N\!\left(S_n, \; \frac{\psi_n^4}{\sigma^2 + \psi_n^2}\right).
\end{equation}
Equations~\eqref{eq:normal_ss}--\eqref{eq:transition} show that the entire sequential problem can be written as a Markov chain on the one-dimensional sufficient statistic $S_n$, with known Gaussian transitions.

\paragraph{Bellman equation.} Let $h_n(s)$ denote the expected loss of the optimal terminal decision at stage $n$ when $S_n = s$:
\begin{equation}\label{eq:terminal}
h_n(s) = \min_{d \in \{d_0, d_1\}} E[R(\theta, d) \mid S_n = s].
\end{equation}
For the loss $R(\theta, d_1) = -\theta$, $R(\theta, d_0) = 0$, this gives $h_n(s) = \min(0, -s)$: stop and choose treatment if $s > 0$, control if $s \leq 0$. Define $V_n(s)$ as the optimal expected loss-to-go from state $s$ at stage $n$, with horizon $T$. The Bellman recursion is
\begin{equation}\label{eq:bellman}
V_n(s) = \min\!\Big\{h_n(s), \;\; c + \int_{-\infty}^{\infty} V_{n+1}(s') \, p(s' \mid S_n = s) \, ds' \Big\},
\end{equation}
with terminal condition $V_T(s) = h_T(s)$. The first argument is the loss from stopping now; the second is the cost of one more patient plus the expected future optimal loss, with the transition density $p(s' \mid S_n = s)$ given by \eqref{eq:transition}. The optimal policy stops at stage $n$ whenever $h_n(s)$ is smaller than the continuation value.

\paragraph{Gridding algorithm.} \citet{brockwell2003} evaluate \eqref{eq:bellman} on a fine grid $\mathcal{G} = \{s_1, \ldots, s_G\}$ over the range of $S_n$. In their clinical trial example, $\mathcal{G}$ spans $[-6, 6]$ with $G = 12{,}000$ points. The algorithm proceeds backward from $T$:
\begin{enumerate}
\item[\textbf{Step 1.}] At $n = T$, set $V_T(s_j) = h_T(s_j)$ for all $s_j \in \mathcal{G}$.
\item[\textbf{Step 2.}] For $n = T-1, T-2, \ldots, 0$: compute the continuation value 
\[
Q_n(s_j) = c + \sum_{s_i \in \mathcal{G}} V_{n+1}(s_i) \, p(s_i \mid S_n = s_j) \, \Delta s
\]
by numerical quadrature over the Gaussian transition \eqref{eq:transition}.
Set 
\[
V_n(s_j) = \min\{h_n(s_j), Q_n(s_j)\}.
\]
\end{enumerate}
Because $\psi_n^2$ is deterministic, the transition density width changes with $n$ but is the same for all $s_j$ at a given $n$, so the quadrature weights can be precomputed once per stage.

\begin{figure}[t]
\centering
\includegraphics[width=0.85\textwidth]{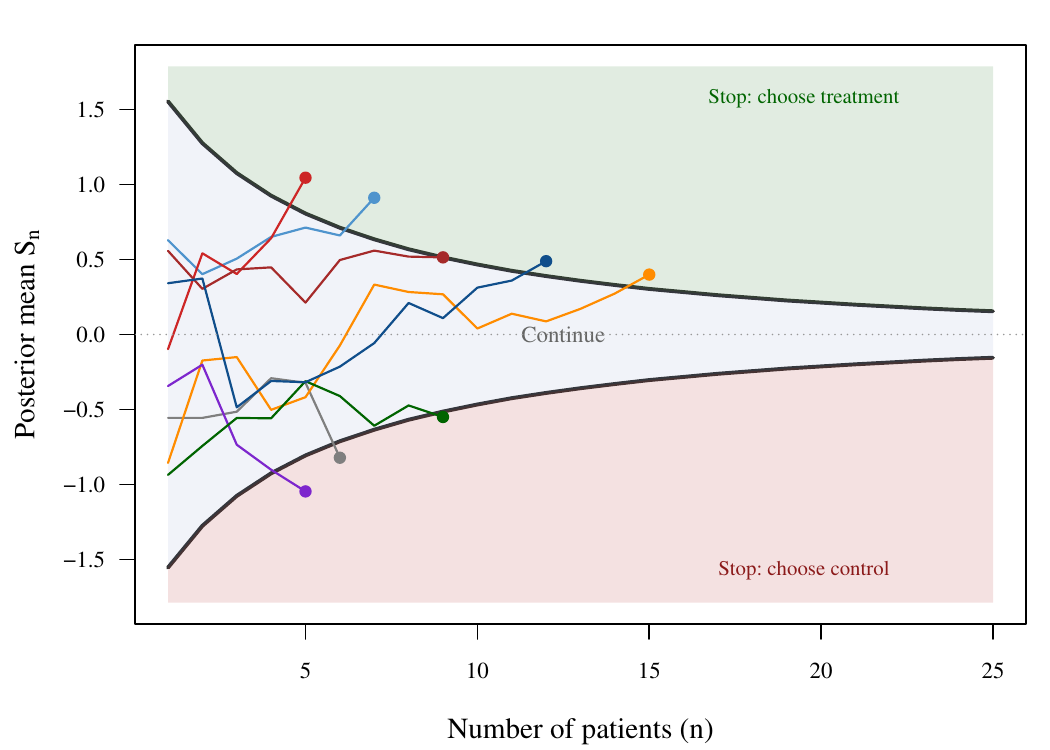}
\caption{Optimal stopping regions from backward induction on the Normal model \eqref{eq:normal_model} with $\sigma^2 = 4$, $\sigma_0^2 = 1$, and sampling cost $c = 0.005$ per patient. The upper region (green) indicates stopping in favour of treatment; the lower region (red) indicates stopping in favour of control; the middle region indicates continuing. Coloured curves show eight simulated trial paths with different true $\theta$ values, each terminating when it crosses a boundary.}
\label{fig:stopping}
\end{figure}

\paragraph{Triangular stopping regions.} The optimal policy partitions the $(n, S_n)$ plane into three regions: stop and choose treatment ($S_n$ large), stop and choose control ($S_n$ small), and continue (intermediate $S_n$). These regions form a triangle that narrows with $n$: early in the trial, the continuation region is wide (much can still be learned); as $n$ approaches $T$, the boundaries converge and the trial is forced to a decision. Figure~\ref{fig:stopping} shows the result for the Normal model with $\sigma^2 = 4$, $\sigma_0^2 = 1$, and sampling cost $c = 0.005$, together with eight simulated trial paths that terminate when they cross a boundary. The shape resembles the triangular continuation region in frequentist group sequential designs, but here the boundaries are derived from the loss function and the prior rather than from alpha-spending constraints.

\paragraph{Extensions.} When $\sigma^2$ is unknown, $\psi_n^2$ is no longer deterministic and must be tracked as a second sufficient statistic; adding $S_n^{(2)} = E[\theta^2 \mid \text{data}]$ restores the Markov property, and the grid becomes two-dimensional. For multi-arm trials, each arm has its own posterior mean, and the grid dimension equals the number of arms. The computational cost grows polynomially in the grid size and linearly in the horizon, making problems with 2--3 arms and horizons of several hundred patients tractable on modern hardware.

The Bayesian sequential framework has a recent frequentist counterpart in $e$-values \citep{grunwald2024}: nonnegative random variables satisfying $E_P[E] \leq 1$ under the null that can be multiplied across sequential analyses while preserving type-I error. Growth-rate-optimal $e$-variables take the form of Bayes factors, providing a bridge between Bayesian and frequentist sequential methods. \citet{sokolova2026} provide a practical comparison of $e$-value, group sequential, and Bayesian monitoring approaches for two-arm binary trials, finding that calibrated group sequential methods achieve the highest power while $e$-values provide robust anytime-valid control at moderate power cost.

\subsection{Binary endpoints: exact backward induction}\label{sec:binary}

The Normal model of the preceding section illustrates backward induction on a continuous sufficient statistic, requiring a discretisation grid and numerical quadrature. We now develop the binary analogue, where the natural integer lattice of success counts replaces the grid entirely, yielding exact optimal stopping rules with no approximation.

\paragraph{Model.} Consider a two-arm trial with binary outcomes and balanced allocation (one patient per arm per stage). Let $p_i$ denote the success probability on arm $i \in \{0, 1\}$, with independent conjugate priors $p_i \sim \text{Beta}(\alpha_i, \beta_i)$. After $k$ stages, arm $i$ has produced $s_i$ successes and $k - s_i$ failures, giving posterior
\begin{equation}\label{eq:binary_post}
p_i \mid s_i, k \sim \text{Beta}(\alpha_i + s_i, \, \beta_i + k - s_i).
\end{equation}

\paragraph{Sufficient statistics.} The state of the trial at stage $k$ is fully described by the pair $(s_1, s_0) \in \{0, \ldots, k\}^2$. A key difference from the Normal model is that the state cannot be reduced to one dimension. In the Normal case, the posterior variance $\psi_n^2$ is deterministic, so the scalar posterior mean $S_n$ is the sole state variable. In the Beta-Binomial case, the predictive transition probabilities
\begin{equation}\label{eq:binary_predict}
\pi_i(s_i, k) = \Pr(Y_{i,k+1} = 1 \mid s_i, k) = \frac{\alpha_i + s_i}{\alpha_i + \beta_i + k}
\end{equation}
depend on $s_1$ and $s_0$ individually, not merely on their difference; two states with the same posterior mean treatment effect $\hat{\delta}_k = \pi_1(s_1,k) - \pi_0(s_0,k)$ but different $(s_1, s_0)$ have different continuation values. The state is therefore irreducibly two-dimensional.

\paragraph{Terminal value.} Using the same loss structure as the Normal model---$R(\mathbf{p}, d_1) = -(p_1 - p_0)$ for choosing treatment, $R(\mathbf{p}, d_0) = 0$ for choosing control---the expected loss of the optimal terminal decision at stage $k$ is
\begin{equation}\label{eq:binary_terminal}
h_k(s_1, s_0) = -\max(0, \, \hat{\delta}_k),
\end{equation}
with $\hat{\delta}_k = (\alpha_1 + s_1)/(\alpha_1 + \beta_1 + k) - (\alpha_0 + s_0)/(\alpha_0 + \beta_0 + k)$.

\paragraph{Bellman equation.} Under balanced allocation, the next stage produces independent outcomes $(Y_1, Y_0) \in \{0,1\}^2$, transitioning from state $(s_1, s_0)$ to $(s_1 + Y_1, s_0 + Y_0)$ with probability $\pi_1^{Y_1}(1 - \pi_1)^{1-Y_1} \cdot \pi_0^{Y_0}(1 - \pi_0)^{1-Y_0}$. The optimal loss-to-go satisfies
\begin{align}
V_k(s_1, s_0) &= \min\!\bigg\{h_k(s_1,s_0), \nonumber\\
& \quad c + \!\sum_{y_1=0}^{1}\sum_{y_0=0}^{1} V_{k+1}(s_1 + y_1, \, s_0 + y_0) \, \Pr(Y_1 = y_1, Y_0 = y_0 \mid s_1, s_0, k)\bigg\}, \label{eq:binary_bellman}
\end{align}
with terminal condition $V_T(s_1, s_0) = h_T(s_1, s_0)$.

Each continuation value is a weighted sum of exactly four successor values, computed in closed form from \eqref{eq:binary_predict}. No numerical quadrature is required---the transition probabilities are exact. This is the principal computational advantage over the Normal model, where the Gaussian transition \eqref{eq:transition} must be integrated numerically over the grid $\mathcal{G}$.

\begin{proposition}[Sufficient-statistic reduction for binary trials]\label{prop:binary}
In a two-arm trial with balanced allocation, Beta$(\alpha_i, \beta_i)$ priors, and per-stage cost $c > 0$, the optimal stopping rule depends on data only through the pair of success counts $(s_1, s_0)$ at each stage $k$. The backward induction \eqref{eq:binary_bellman} has $(k+1)^2$ states at stage $k$, for a total of $\sum_{k=0}^{T}(k+1)^2 = O(T^3)$ evaluations, each requiring $O(1)$ arithmetic operations.
\end{proposition}

The proof is immediate from the sufficiency of $(s_i, k)$ for the Beta model \eqref{eq:binary_post} and the closed-form predictive \eqref{eq:binary_predict}. For a trial with $T = 200$ stages (400 patients), the total computation is approximately $2.7 \times 10^6$ evaluations---completed in under one second on standard hardware. By contrast, the Normal model with $G = 4000$ grid points and horizon $T = 50$ requires $2 \times 10^5$ evaluations but incurs discretisation error from the grid spacing $\Delta s$.

\begin{figure}[t]
\centering
\includegraphics[width=0.85\textwidth]{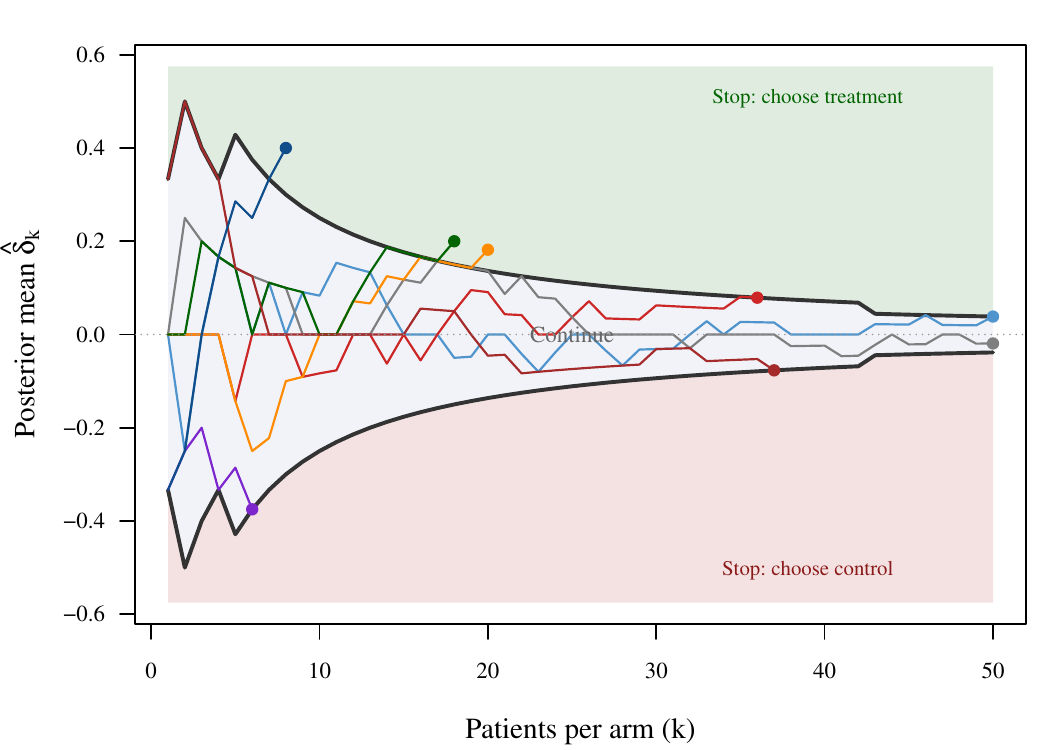}
\caption{Optimal stopping regions from backward induction on the two-arm Beta-Binomial model \eqref{eq:binary_post} with $\text{Beta}(1,1)$ priors and sampling cost $c = 0.0005$ per stage. The regions are projected onto $\hat{\delta}_k$, the posterior mean treatment effect, for direct comparison with the Normal model (Figure~\ref{fig:stopping}). Boundaries are discrete, reflecting the integer lattice of success counts. Coloured curves show eight simulated trial paths with $p_0 = 0.30$ and various $p_1$ values.}
\label{fig:binary_stopping}
\end{figure}

\paragraph{Stopping regions.} Figure~\ref{fig:binary_stopping} shows the optimal stopping regions for the binary model with $\text{Beta}(1,1)$ priors and sampling cost $c = 0.0005$ per stage. The regions are projected onto the posterior mean treatment effect $\hat{\delta}_k$ for comparison with Figure~\ref{fig:stopping}. The qualitative structure---a triangular continuation region that narrows with $k$---is the same as in the Normal model, but the boundaries are discrete (reflecting the integer lattice of success counts) rather than smooth.

\paragraph{Calibrated backward induction.} The terminal value \eqref{eq:binary_terminal} rewards any positive treatment effect regardless of the strength of evidence, leading to early stopping that is optimal for expected utility but yields low power when combined with a stringent declaration threshold. An alternative embeds the evidence requirement directly in the utility:
\begin{equation}\label{eq:calibrated_terminal}
h_k^{(\gamma)}(s_1, s_0) = -\max\!\big(0,\; \hat{\delta}_k \cdot \mathbf{1}\{\Pr(p_1 > p_0 \mid s_1, s_0, k) > \gamma\}\big),
\end{equation}
where $\gamma$ is the posterior probability threshold for declaring treatment superior. This ``calibrated'' terminal value is zero unless the treatment effect estimate $\hat{\delta}_k$ is positive \emph{and} the posterior evidence exceeds~$\gamma$. The uncalibrated version \eqref{eq:binary_terminal} is the special case $\gamma = 0$. The posterior probability $\Pr(p_1 > p_0 \mid s_1, s_0, k)$ is computed via a normal approximation to the difference of independent Beta posteriors, preserving the $O(T^3)$ computational complexity.

By making the backward induction aware that declarations require strong evidence, the calibrated policy samples longer than its uncalibrated counterpart: it recognises that stopping early has no value if the accumulated data cannot support a declaration. The simulation study in Section~\ref{sec:simulation} compares both variants.

\subsection{P\'olya-Gamma bridge to logistic regression}\label{sec:pgbridge}

The Beta-Binomial backward induction of Section~\ref{sec:binary} requires no covariate adjustment: the success probability $p_i$ on each arm is exchangeable across patients. When patient-level covariates $\mathbf{x}_j \in \mathbb{R}^p$ are available---baseline risk factors, biomarkers, or dose levels---the model becomes logistic:
\[
Y_j \mid \boldsymbol{\beta} \sim \text{Bernoulli}\!\left(\text{logit}^{-1}(\mathbf{x}_j'\boldsymbol{\beta})\right), \qquad \boldsymbol{\beta} \sim N(\mathbf{b}_0, \mathbf{B}_0),
\]
and the Beta conjugacy of Section~\ref{sec:binary} is lost. The P\'olya-Gamma augmentation of \citet{polson2013}, introduced in Section~\ref{sec:priors}, restores a conditional conjugacy that connects back to the Normal model.

\begin{proposition}[P\'olya-Gamma bridge]\label{prop:pgbridge}
Under the P\'olya-Gamma augmentation \eqref{eq:pg}, conditional on latent variables $\omega_j \sim \mathrm{PG}(1, \mathbf{x}_j'\boldsymbol{\beta})$, the posterior of the regression coefficients is Gaussian:
\begin{equation}\label{eq:pg_posterior}
\boldsymbol{\beta} \mid \mathbf{Y}, \boldsymbol{\omega} \sim N(\mathbf{m}_n, \mathbf{V}_n), \qquad \mathbf{V}_n = (\mathbf{B}_0^{-1} + \mathbf{X}'\boldsymbol{\Omega}\mathbf{X})^{-1}, \quad \mathbf{m}_n = \mathbf{V}_n(\mathbf{B}_0^{-1}\mathbf{b}_0 + \mathbf{X}'\boldsymbol{\kappa}),
\end{equation}
where $\boldsymbol{\Omega} = \mathrm{diag}(\omega_1, \ldots, \omega_n)$ and $\kappa_j = y_j - 1/2$. Replacing $\boldsymbol{\omega}$ by its conditional expectation $\hat{\omega}_j = E[\omega_j \mid \boldsymbol{\beta} = \hat{\boldsymbol{\beta}}] = \tanh(\mathbf{x}_j'\hat{\boldsymbol{\beta}}/2)/(2\mathbf{x}_j'\hat{\boldsymbol{\beta}})$ at the posterior mode $\hat{\boldsymbol{\beta}}$ yields a Gaussian approximation whose precision matrix $\mathbf{B}_0^{-1} + \mathbf{X}'\hat{\boldsymbol{\Omega}}\mathbf{X}$ reduces backward induction to the Normal model framework of Section~\ref{sec:sequential} on a $p$-dimensional grid.
\end{proposition}

The conditional Gaussian form \eqref{eq:pg_posterior} follows directly from the P\'olya-Gamma identity \eqref{eq:pg} applied to each binomial likelihood factor; see \citet{polson2013}, Theorem~1. The plug-in step replaces the latent $\boldsymbol{\omega}$ by a deterministic function of the posterior mode $\hat{\boldsymbol{\beta}}$, yielding a Laplace-type approximation to the marginal posterior. This approximation is accurate when the posterior is concentrated (large samples) but may degrade with small samples or complete separation; see the limitations discussion in Section~\ref{sec:discussion}.

This connects three sections of the paper into a single pipeline. The conditional means prior of \citet{bedrick1996} (Section~\ref{sec:priors}) translates clinical beliefs about response rates into a prior $\pi(\boldsymbol{\beta})$ via the link function. For dose-response models, elicitation is often simplest when expressed through the LD50 (dose at 50\% response), which satisfies $\beta_0 + \beta_1 \cdot \text{LD50} = 0$ under the logistic link. A prior on the LD50 and the slope parameter (related to $\beta_1$ through the derivative of the logistic function) induces a joint prior on $(\beta_0, \beta_1)$ via the reparameterisation $\beta_0 = -\beta_1 \cdot \text{LD50}$. This induced prior then enters the P\'olya-Gamma scheme, which provides posterior samples via a tuning-free Gibbs sampler alternating between the latent $\boldsymbol{\omega}$ and the Gaussian conditional \eqref{eq:pg_posterior}, enabling real-time Bayesian monitoring during the trial with no Metropolis--Hastings step. The Gaussian conditional structure further enables backward induction on a grid of posterior mean vectors, with transitions that are approximately Gaussian and precision matrices that are approximately determined by the design. For a two-parameter model (intercept and treatment effect, $p = 2$), the resulting grid is two-dimensional---the same dimensionality as the exact Beta-Binomial lattice, but now expressed on the logistic scale so that additional covariates can in principle be incorporated by enlarging $\boldsymbol{\beta}$. This hybrid workflow---interpretable prior elicitation through conditional means, efficient computation through P\'olya-Gamma augmentation, and optimal sequential decisions through backward induction---illustrates how the three components combine, though extension to higher-dimensional $\boldsymbol{\beta}$ would require sparse-grid or dimension-reduction techniques to keep the backward induction tractable.

\paragraph{Validation.} Table~\ref{tab:pg_validation} quantifies the accuracy of the P\'olya-Gamma Laplace approximation for the posterior probability of treatment superiority $\Pr(p_1 > p_0 \mid \text{data})$, using a logistic-scale prior moment-matched to the same $\text{Beta}(1,1)$ hyperparameters via the digamma and trigamma functions, with the full prior covariance between intercept and treatment effect induced by the change of variables. Across all sample sizes tested ($n = 10$ to $200$ per arm), the mean absolute error across random datasets remains below 0.01. The approximation is most accurate when the treatment effect is large (posterior concentrated away from 0.5); under the null, errors are slightly larger but remain small, with the worst-case mean of 0.010 occurring at $n = 50$. These results support the use of the P\'olya-Gamma bridge for moderately sized trials, with the caveat that small-sample applications should verify against exact Beta posterior calculations.

\begin{table}[t]
\centering
\caption{Accuracy of the P\'olya-Gamma Laplace approximation to $\Pr(p_1 > p_0 \mid \text{data})$. Exact values are computed from independent $\text{Beta}(1 + s_j,\, 1 + n - s_j)$ posteriors via Monte Carlo ($10^5$ draws). The PG-Laplace column uses a logistic regression parameterisation with iterative P\'olya-Gamma plug-in at the posterior mode, with a logistic-scale prior moment-matched to the same $\text{Beta}(1,1)$ hyperparameters, including the off-diagonal covariance between intercept and treatment effect induced by the change of variables.}
\label{tab:pg_validation}
\begin{tabular}{rrrrrr}
\hline
$n$ & $\delta$ & $s_1/n$ & $s_0/n$ & Exact & PG-Laplace \\
\hline
$5$ & $0.00$ & $0.40$ & $0.40$ & $0.502$ & $0.500$ \\
 & $0.05$ & $0.40$ & $0.40$ & $0.499$ & $0.500$ \\
 & $0.15$ & $0.40$ & $0.40$ & $0.500$ & $0.500$ \\
 & $0.25$ & $0.60$ & $0.40$ & $0.718$ & $0.715$ \\
[3pt]
$10$ & $0.00$ & $0.30$ & $0.30$ & $0.498$ & $0.500$ \\
 & $0.05$ & $0.40$ & $0.30$ & $0.670$ & $0.672$ \\
 & $0.15$ & $0.40$ & $0.30$ & $0.671$ & $0.672$ \\
 & $0.25$ & $0.60$ & $0.30$ & $0.900$ & $0.901$ \\
[3pt]
$20$ & $0.00$ & $0.30$ & $0.30$ & $0.501$ & $0.500$ \\
 & $0.05$ & $0.35$ & $0.30$ & $0.627$ & $0.632$ \\
 & $0.15$ & $0.45$ & $0.30$ & $0.828$ & $0.833$ \\
 & $0.25$ & $0.55$ & $0.30$ & $0.940$ & $0.942$ \\
[3pt]
$50$ & $0.00$ & $0.30$ & $0.30$ & $0.500$ & $0.500$ \\
 & $0.05$ & $0.36$ & $0.30$ & $0.735$ & $0.743$ \\
 & $0.15$ & $0.44$ & $0.30$ & $0.926$ & $0.929$ \\
 & $0.25$ & $0.56$ & $0.30$ & $0.995$ & $0.996$ \\
[3pt]
$100$ & $0.00$ & $0.30$ & $0.30$ & $0.499$ & $0.500$ \\
 & $0.05$ & $0.35$ & $0.30$ & $0.773$ & $0.783$ \\
 & $0.15$ & $0.45$ & $0.30$ & $0.986$ & $0.987$ \\
 & $0.25$ & $0.55$ & $0.30$ & $1.000$ & $1.000$ \\
[3pt]
$200$ & $0.00$ & $0.30$ & $0.30$ & $0.500$ & $0.500$ \\
 & $0.05$ & $0.35$ & $0.30$ & $0.857$ & $0.867$ \\
 & $0.15$ & $0.45$ & $0.30$ & $0.999$ & $0.999$ \\
 & $0.25$ & $0.55$ & $0.30$ & $1.000$ & $1.000$ \\
[3pt]
\hline
\multicolumn{6}{l}{\small Monte Carlo validation (50 random datasets):} \\
\hline
$n$ & $\delta$ & \multicolumn{2}{c}{Mean abs.\ error} & \multicolumn{2}{c}{Max abs.\ error} \\
\hline
$10$ & $0.00$ & \multicolumn{2}{c}{$0.0085$} & \multicolumn{2}{c}{$0.0281$} \\
$10$ & $0.15$ & \multicolumn{2}{c}{$0.0038$} & \multicolumn{2}{c}{$0.0182$} \\
$10$ & $0.25$ & \multicolumn{2}{c}{$0.0019$} & \multicolumn{2}{c}{$0.0115$} \\
$50$ & $0.00$ & \multicolumn{2}{c}{$0.0097$} & \multicolumn{2}{c}{$0.0303$} \\
$50$ & $0.15$ & \multicolumn{2}{c}{$0.0038$} & \multicolumn{2}{c}{$0.0097$} \\
$50$ & $0.25$ & \multicolumn{2}{c}{$0.0012$} & \multicolumn{2}{c}{$0.0079$} \\
$200$ & $0.00$ & \multicolumn{2}{c}{$0.0093$} & \multicolumn{2}{c}{$0.0213$} \\
$200$ & $0.15$ & \multicolumn{2}{c}{$0.0009$} & \multicolumn{2}{c}{$0.0084$} \\
$200$ & $0.25$ & \multicolumn{2}{c}{$0.0000$} & \multicolumn{2}{c}{$0.0007$} \\
\hline
\end{tabular}
\end{table}

\subsection{Simulation study}\label{sec:simulation}

To illustrate the operating characteristics of binary backward induction relative to established methods, we simulate a two-arm trial with control success rate $p_0 = 0.30$ under four treatment effect scenarios: $\delta = p_1 - p_0 \in \{0, 0.05, 0.15, 0.25\}$. The predictive probability, group sequential, and fixed-sample designs use a maximum of $N = 100$ patients per arm; backward induction uses a horizon of $T = 200$ stages, though in practice it stops far earlier.

We compare four designs: (i)~backward induction from Section~\ref{sec:binary} with $\text{Beta}(1,1)$ priors, per-stage cost $c = 0.0005$, declaring treatment superior only when $\Pr(p_1 > p_0 \mid \text{data}) > 0.975$ (the Bellman recursion is exact on the integer lattice using posterior mean differences; at the stopping time, the declaration uses Monte Carlo draws from the exact Beta posteriors); (ii)~predictive probability monitoring \citep{marion2025} with efficacy/futility thresholds $0.95/0.05$ and 10 equally spaced interim looks; (iii)~O'Brien--Fleming group sequential design \citep{obrienfleming1979} with 5 one-sided looks using the standard boundary approximation $z_j^* = z_\alpha \sqrt{K/j}$; and (iv)~a fixed-sample design with a one-sided $z$-test at level 0.025.

\begin{table}[t]
\centering
\caption{Operating characteristics of four designs for a two-arm binary trial with $p_0 = 0.30$ and balanced allocation ($k$ patients per arm). $E[N]$: expected patients per arm; Power: probability of declaring treatment superior (type~I error under the null). Backward induction uses $\text{Beta}(1,1)$ priors, per-stage cost $c = 0.0005$, horizon $T = 200$, and declares treatment only when $\Pr(p_1 > p_0 \mid \text{data}) > 0.975$. Predictive probability uses \citet{marion2025} with efficacy/futility thresholds $0.95/0.05$ and 10 equally spaced looks. O'Brien--Fleming uses 5 looks with one-sided boundaries. Predictive probability, O'Brien--Fleming, and fixed-sample designs use $N = 100$ patients per arm.}
\label{tab:simulation}
\begin{tabular}{llrr}
\hline
$\delta$ & Method & $E[N]$ & Power \\
\hline
$0.00$ & Backward induction & $25.9$ & $0.029$ \\
 & Predictive prob. & $42.4$ & $0.043$ \\
 & O'Brien--Fleming & $99.5$ & $0.033$ \\
 & Fixed-sample & $100.0$ & $0.028$ \\
[3pt]
$0.05$ & Backward induction & $25.5$ & $0.047$ \\
 & Predictive prob. & $52.9$ & $0.145$ \\
 & O'Brien--Fleming & $97.8$ & $0.134$ \\
 & Fixed-sample & $100.0$ & $0.120$ \\
[3pt]
$0.15$ & Backward induction & $20.0$ & $0.100$ \\
 & Predictive prob. & $62.8$ & $0.587$ \\
 & O'Brien--Fleming & $85.4$ & $0.625$ \\
 & Fixed-sample & $100.0$ & $0.600$ \\
[3pt]
$0.25$ & Backward induction & $14.3$ & $0.179$ \\
 & Predictive prob. & $47.5$ & $0.931$ \\
 & O'Brien--Fleming & $64.8$ & $0.959$ \\
 & Fixed-sample & $100.0$ & $0.954$ \\
\hline
\end{tabular}
\end{table}

Table~\ref{tab:simulation} reports expected sample size per arm ($E[N]$) and the probability of declaring treatment superior (power when $\delta > 0$; type~I error when $\delta = 0$). Under the null, backward induction and the fixed-sample design maintain type~I error near 0.025; the O'Brien--Fleming approximation is slightly liberal at 0.033; predictive probability monitoring is more inflated at 0.043, reflecting the difficulty of calibrating Bayesian stopping thresholds to frequentist error rates. The backward induction design achieves dramatically lower expected sample size---14 to 26 patients per arm versus 42 to 100 for the other methods---but at the cost of substantially lower power: 0.179 versus 0.954 for the fixed-sample design at $\delta = 0.25$. (All reported proportions are based on $10\,000$ simulations, except the calibrated variant which uses $5\,000$; Monte Carlo standard errors are below 0.005 throughout.)

This trade-off reflects a fundamental tension between optimal stopping and frequentist-calibrated declaration. The backward induction stops early because the loss function values patient savings: once the expected benefit of further sampling falls below the per-stage cost $c$, continued enrolment is suboptimal. But early stopping with few patients means the posterior probability $\Pr(p_1 > p_0 \mid \text{data})$ rarely exceeds the stringent declaration threshold $\gamma = 0.975$. The design is not optimising for power; it is optimising for expected utility, which trades off the information from additional patients against the cost of enrolling them. In settings where the primary objective is to minimise patient exposure to an inferior treatment rather than to maximise the probability of a positive declaration---for example, paediatric or rare-disease trials where each patient has high individual value---this trade-off may be acceptable or even preferred.

The predictive probability and group sequential methods occupy a middle ground: they allow early stopping for overwhelming evidence (at $\delta = 0.25$, expected sample sizes fall to 48 and 65 patients per arm respectively, compared to 100 for the fixed-sample design) while maintaining power close to the fixed-sample benchmark. These comparisons are illustrative rather than definitive: the methods differ in their calibration targets (utility versus type~I error), horizons, and number of interim looks, so no single comparison can rank them globally. The value of the backward induction framework is not that it dominates these alternatives, but that it provides the exact optimal policy for a stated utility function, making the trade-offs between sample size, power, and error rate transparent rather than implicit.

\begin{figure}[t]
\centering
\includegraphics[width=0.95\textwidth]{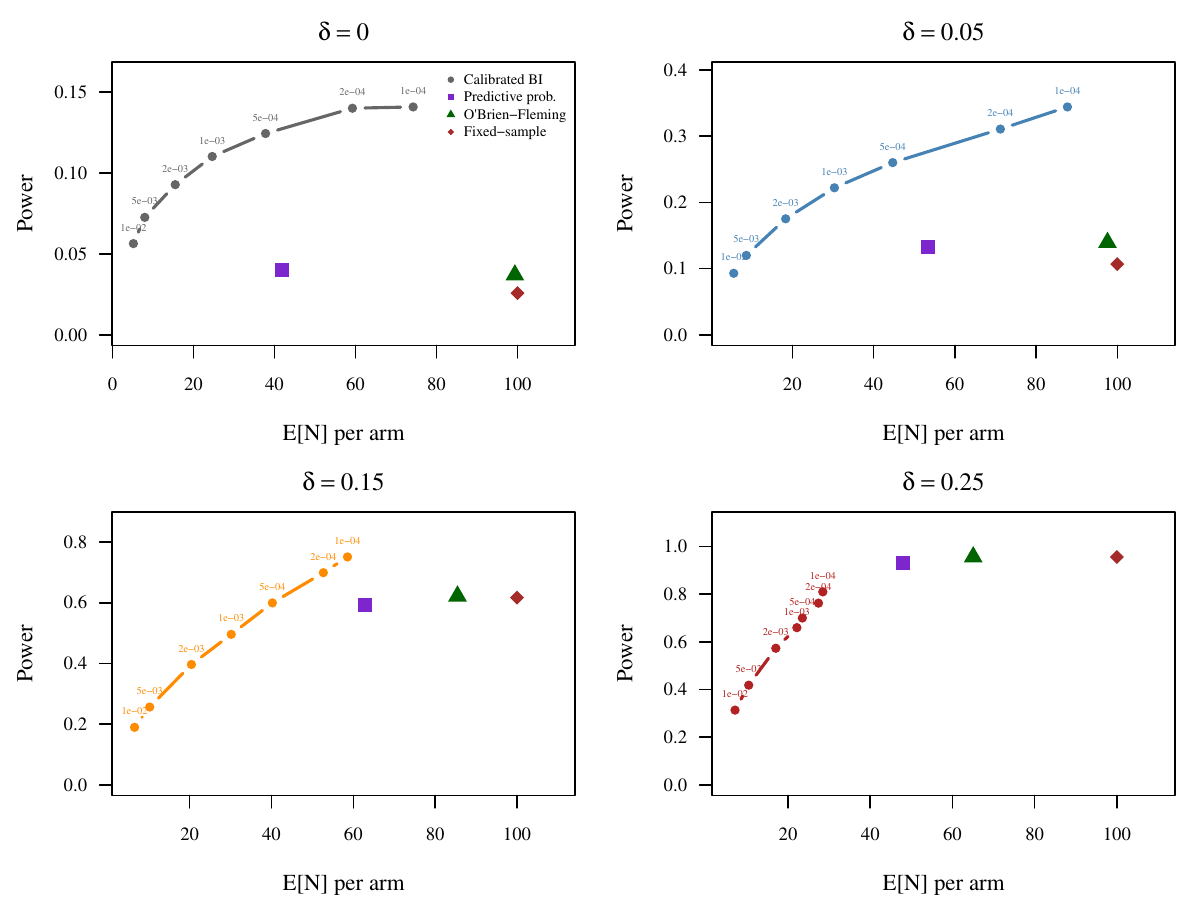}
\caption{Power frontier for calibrated backward induction (connected circles) across per-stage costs $c \in \{10^{-4}, \ldots, 10^{-2}\}$, with competitors shown as fixed symbols. Each panel corresponds to a treatment effect scenario ($\delta = p_1 - p_0$). Lower cost yields larger expected sample size and higher power. The frontier illustrates the trade-off available to the clinician: the per-stage cost $c$ controls the operating point.}
\label{fig:power_frontier}
\end{figure}

\paragraph{Calibrated backward induction and the power frontier.} Embedding the declaration threshold in the terminal utility \eqref{eq:calibrated_terminal} substantially changes the operating characteristics. Figure~\ref{fig:power_frontier} shows the power frontier traced by varying the per-stage cost $c$ from $10^{-4}$ to $10^{-2}$, with the calibrated backward induction using $\gamma = 0.975$ and $T = 200$. At the same cost $c = 5 \times 10^{-4}$ used for the uncalibrated design, calibration raises power at $\delta = 0.25$ from 0.18 to 0.70 while $E[N]$ increases from 14 to 24 patients per arm---a doubling, but still well below the 48--100 required by competitors. Reducing the cost further to $c = 10^{-4}$ pushes power to 0.81 with $E[N] = 29$. Both calibrated designs inflate type~I error (0.12--0.14 versus 0.03 for the uncalibrated rule). The inflated type~I error arises because backward induction optimises expected utility, not frequentist error rates: the calibrated policy samples longer, and the sequential nature means the posterior probability has more opportunities to exceed $\gamma$ under the null.

The power frontier makes the trade-off explicit. At low cost (large trials), the calibrated design approaches the power of group sequential and predictive probability methods (0.81 versus 0.93--0.96 at $\delta = 0.25$), though a gap remains because the Bayesian policy does not optimise frequentist power directly; at high cost (small trials), it achieves large sample savings but with reduced power. The clinician selects the operating point appropriate to the clinical context by choosing~$c$.

\begin{table}[t]
\centering
\caption{Prior sensitivity of calibrated backward induction for a two-arm binary trial with $p_0 = 0.30$. All designs use $T = 200$, $c = 0.0005$, $\gamma = 0.975$. Symmetric priors are placed on both arms.}
\label{tab:prior_sensitivity}
\begin{tabular}{llrr}
\hline
$\delta$ & Prior & $E[N]$ & Power \\
\hline
$0.00$ & Beta(1,1) & $36.7$ & $0.135$ \\
 & Beta(0.5,0.5) & $36.1$ & $0.156$ \\
 & Beta(3,7) & $38.4$ & $0.078$ \\
[3pt]
$0.05$ & Beta(1,1) & $44.9$ & $0.267$ \\
 & Beta(0.5,0.5) & $42.8$ & $0.306$ \\
 & Beta(3,7) & $49.0$ & $0.204$ \\
[3pt]
$0.15$ & Beta(1,1) & $40.7$ & $0.588$ \\
 & Beta(0.5,0.5) & $39.6$ & $0.650$ \\
 & Beta(3,7) & $48.3$ & $0.583$ \\
[3pt]
$0.25$ & Beta(1,1) & $23.8$ & $0.704$ \\
 & Beta(0.5,0.5) & $22.5$ & $0.759$ \\
 & Beta(3,7) & $32.3$ & $0.742$ \\
[3pt]
\hline
\end{tabular}
\end{table}

Table~\ref{tab:prior_sensitivity} shows the sensitivity of the calibrated design to the choice of prior. The informative $\text{Beta}(3,7)$ prior, centred at the null control rate $p_0 = 0.30$, reduces type~I error from 0.14 to 0.08 while maintaining power at $\delta = 0.25$ of 0.74, illustrating how shrinkage toward the prior mean on both arms can partially mitigate the type~I error inflation inherent in sequential Bayesian designs. The $\text{Beta}(0.5, 0.5)$ Jeffreys prior slightly increases both power and type~I error relative to the uniform prior, consistent with its heavier tails concentrating posterior mass near 0 and~1.

\section{Decision-Theoretic Trial Design}\label{sec:decision}

A clinical trial involves a sequence of decisions: which dose to administer, which patients to enrol, when to stop, and what to conclude. The frequentist framework treats some of these as design parameters fixed before the trial and others as inference conducted after it. The Bayesian decision-theoretic framework treats them uniformly as decisions under uncertainty, optimised with respect to a utility function.

\subsection{Decisions under uncertainty}

\citet{lindley1985} articulated the principle that any choice among alternatives whose consequences depend on unknown quantities is a decision problem, requiring a probability distribution over the unknowns and a utility function over the consequences. In a clinical trial, the unknowns include treatment effects, dose-response curves, and subgroup-specific outcomes; the consequences include patient survival, toxicity, regulatory approval, and cost. Every design choice---randomisation ratio, stopping boundary, dose escalation rule---is implicitly a decision function mapping accumulating data to actions, and coherence demands that this function maximise expected utility.

This observation connects the backward induction of Section~\ref{sec:sequential} to the designs we develop here. The gridding algorithm of \citet{brockwell2003} already solves a decision problem: the utility in \eqref{eq:utility} trades off the reward from a correct terminal decision against the cost of continued sampling. The designs below extend this idea to richer utility functions that incorporate efficacy, toxicity, and information gain simultaneously.

The value of information provides the key link between decision theory and trial design. Given current data $D_n$, the expected value of sample information (EVSI) from enrolling the next cohort is the difference between the expected utility of the optimal decision after observing the new data and the expected utility of deciding now. When EVSI falls below the per-patient cost, sampling should stop. The predictive probability framework of \citet{marion2025} can be viewed as an approximation to this calculation: a low predictive probability of trial success implies that the information still to be gained is unlikely to change the final decision.

\subsection{Utility-based dose-finding: the EffTox design}

Traditional phase~I designs seek only the maximum tolerated dose (MTD), treating toxicity as the sole endpoint. This is a poor decision rule when the MTD is not the most effective dose---a situation that arises frequently in oncology, where efficacy may plateau or decline at high doses due to off-target effects.

\citet{thall1998} and \citet{thall2004} developed designs that jointly model efficacy and toxicity, selecting doses to maximise a clinical utility function. Let $\pi_E(x)$ and $\pi_T(x)$ denote the probabilities of efficacy and toxicity at dose $x$. The marginal models use logistic regression:
\begin{align}
\text{logit}[\pi_E(x)] &= \mu_E + \beta_{E,1} x + \beta_{E,2} x^2, \label{eq:efftox_eff} \\
\text{logit}[\pi_T(x)] &= \mu_T + \beta_T x. \label{eq:efftox_tox}
\end{align}
The quadratic term in \eqref{eq:efftox_eff} allows for a non-monotone dose-efficacy relationship, while toxicity is assumed monotone increasing. The joint distribution of (efficacy, toxicity) for each patient is specified through a Gumbel copula with association parameter $\psi$, allowing the two outcomes to be correlated---as they typically are, since drugs that are more biologically active tend to produce both more responses and more side effects.

A utility function $U(\pi_E, \pi_T)$ is elicited from clinicians by specifying the value of each possible outcome combination: efficacy alone ($U(1,0)$), toxicity alone ($U(0,1)$), both ($U(1,1)$), and neither ($U(0,0)$).
Normalising so that $U(1,0) = 100$ and $U(0,1) = 0$, the clinician provides $U(1,1)$ (the penalty for toxicity when the treatment works) and $U(0,0)$ (the value of avoiding both benefit and harm).
These four points, together with indifference contours in the $(\pi_E, \pi_T)$ plane---combinations the clinician judges equally desirable---define a complete utility surface.
The conditional means prior framework of \citet{bedrick1996} discussed in Section~\ref{sec:priors} provides a natural way to specify priors on the dose-response parameters in \eqref{eq:efftox_eff}--\eqref{eq:efftox_tox}: the clinician states prior beliefs about response rates at two or three clinically meaningful doses, and the link function induces priors on~$(\mu_E, \beta_{E,1}, \beta_{E,2}, \mu_T, \beta_T)$.

At each stage, the design:
\begin{enumerate}
\item[(a)] updates the posterior distribution of all model parameters given accumulated data;
\item[(b)] identifies \textit{admissible} doses: those satisfying $\Pr(\pi_E(x) > A_E \mid \text{data}) > p_E$ and $\Pr(\pi_T(x) < A_T \mid \text{data}) > p_T$, where $A_E$ and $A_T$ are clinically meaningful efficacy and toxicity thresholds;
\item[(c)] among admissible doses, selects the one maximising posterior expected utility\\$E[U(\pi_E(x), \pi_T(x)) \mid \text{data}]$.
\end{enumerate}
The admissibility constraints act as safety guardrails: no dose with unacceptably low efficacy or high toxicity is ever selected, regardless of its utility. Within the admissible set, utility maximisation balances the two objectives. A comprehensive treatment of the design, including calibration of the threshold probabilities $p_E$ and $p_T$ through simulation, is given by \citet{thall2016}.

The lecanemab trial for early Alzheimer's disease illustrates utility-based dose-finding in practice. \citet{berry2023lecanemab} used a Bayesian phase~2b design with five dose-schedule regimens and placebo. The primary endpoint was change in amyloid PET, but the dose-selection rule incorporated both efficacy (amyloid reduction) and tolerability (amyloid-related imaging abnormalities), producing a dose recommendation that balanced benefit against a clinically significant side effect. The selected regimen proceeded to a confirmatory phase~3 trial that led to FDA approval.

\citet{thall2024} extends EffTox to precision medicine, where utility functions and dose-finding rules are tailored to prognostic subgroups defined by biomarkers. Rather than seeking a single optimal dose for all patients, the design maintains subgroup-specific posterior distributions and selects different doses for different biomarker profiles, reflecting the clinical reality that the optimal efficacy-toxicity trade-off varies across patient populations.

\subsection{Adaptive enrichment}

When a treatment benefits only a biomarker-defined subgroup, enrolling unselected patients wastes resources and exposes non-responders to risk. An enrichment design modifies enrolment criteria based on interim data, starting with broad enrolment and restricting to the responsive subgroup as evidence accumulates.

Formally, consider a trial with $G$ pre-specified subgroups $\mathcal{S}_1, \ldots, \mathcal{S}_G$ defined by a biomarker. At each interim analysis after $n$ patients, the design computes subgroup-specific posterior probabilities $\Pr(\delta_g > 0 \mid D_n)$ for each subgroup $g$, where $\delta_g$ is the treatment effect in subgroup $g$. The decision rule maps these probabilities to one of three actions: continue enrolling all subgroups, restrict enrolment to subgroups with $\Pr(\delta_g > 0 \mid D_n) > \gamma_{\text{enrich}}$, or stop the trial entirely if no subgroup shows sufficient promise. The thresholds $\gamma_{\text{enrich}}$ and the stopping boundaries are calibrated by simulation to control the overall type~I error rate while maximising power in the subgroup that benefits.

The decision at each interim---whether to continue broad enrolment, restrict, or stop---is a multi-armed bandit problem where the ``arms'' are enrolment strategies rather than treatments. This is precisely the structure addressed by the backward induction framework of Section~\ref{sec:sequential}: the state space is the collection of subgroup-specific sufficient statistics, and the utility function penalises both the cost of additional enrolment and the regret from treating non-responders.

The I-SPY~2 platform trial for neoadjuvant breast cancer provides a large-scale example. Patients are assigned to one of eight biomarker-defined subtypes, and each experimental arm is evaluated across up to ten molecular signatures (combinations of subtypes corresponding to possible clinical indications); arms that show $\geq 85\%$ Bayesian predictive probability of success in a confirmatory trial ``graduate,'' while arms that cross a futility boundary are dropped \citep{barker2009, berry2025}. This adaptive enrichment within a platform design has graduated nine of 23 experimental arms (39\%) to phase~III, with four of the nine graduates subsequently receiving marketing approval \citep{berry2025}.

\section{Case Studies}\label{sec:cases}

\subsection{ECMO: the role of priors with minimal data}

The 1985 trial of extracorporeal membrane oxygenation (ECMO) for newborns with persistent pulmonary hypertension \citep{bartlett1985} used a randomised play-the-winner design \citep{wei1978}: adaptive randomisation that favours the more successful treatment. The first patient was randomised to ECMO and survived; the second to conventional medical therapy (CMT) and died; the remaining ten were all randomised to ECMO and all survived. Final tally: 11/11 ECMO survivors, 0/1 CMT survivors.

\citet{ware1989} discussed the statistical challenges; \citet{kass1989} provided the Bayesian perspective. With a uniform prior $\text{Beta}(1,1)$ on each treatment's survival probability, the posteriors are:
\begin{align}
\theta_{\text{ECMO}} \mid \text{data} &\sim \text{Beta}(12, 1), \quad E[\theta_{\text{ECMO}} \mid \text{data}] = 12/13 \approx 0.923, \label{eq:ecmo_post1} \\
\theta_{\text{CMT}} \mid \text{data} &\sim \text{Beta}(1, 2), \quad E[\theta_{\text{CMT}} \mid \text{data}] = 1/3 \approx 0.333. \label{eq:ecmo_post2}
\end{align}
The posterior probability $P(\theta_{\text{ECMO}} > \theta_{\text{CMT}} \mid \text{data})$ can be computed exactly via Thompson's formula \eqref{eq:thompson} and equals $90/91 \approx 0.989$.

The critical issue is sensitivity to the prior. With only one control patient, the posterior for $\theta_{\text{CMT}}$ is dominated by the prior. Historical data suggested approximately 20\% survival with CMT from large observational series. Encoding this as $\text{Beta}(4, 16)$---with prior mean $4/20 = 0.20$ and effective sample size on the order of 20---yields $\theta_{\text{CMT}} \mid \text{data} \sim \text{Beta}(4, 17)$ with posterior mean 0.19, and $P(\theta_{\text{ECMO}} > \theta_{\text{CMT}} \mid \text{data})$ remains above 0.99. The conclusion is robust because the ECMO evidence (11/11) is overwhelming; the prior matters only for the magnitude of the estimated difference, not the direction.

\citet{berry1989ecmo} emphasised that the Bayesian framework naturally handles the controversial stopping rule: under the likelihood principle, inference depends on the data actually observed, not on the stopping rule that generated them. A p-value for the same data depends on whether the investigator planned to stop after 12 patients, planned to stop after one CMT death, or had no fixed plan---the Bayesian posterior does not.

\begin{figure}[t]
\centering
\includegraphics[width=0.85\textwidth]{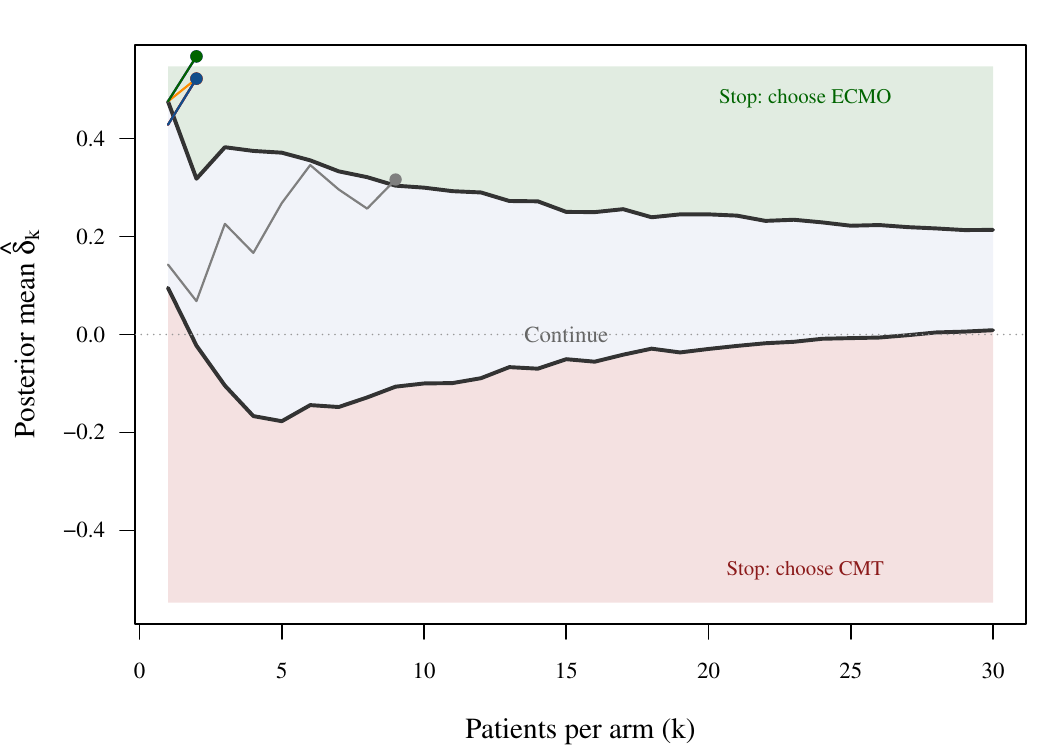}
\caption{Optimal stopping regions from calibrated backward induction for the ECMO trial design, with $\text{Beta}(1,1)$ prior for ECMO and $\text{Beta}(4,16)$ prior for CMT. The asymmetric priors produce a wider stopping region for ECMO superiority (upper) than for CMT superiority (lower). Coloured curves show eight simulated trial paths under the ECMO-like scenario ($p_{\text{ECMO}} = 0.80$, $p_{\text{CMT}} = 0.20$); all terminate within 10 stages.}
\label{fig:ecmo_bi}
\end{figure}

The binary backward induction of Section~\ref{sec:binary} provides a prospective design framework for trials like ECMO. Using a $\text{Beta}(4,16)$ prior for CMT (encoding the historical 20\% survival rate with effective sample size~20) and a $\text{Beta}(1,1)$ prior for ECMO (noninformative for the experimental arm), we compute the calibrated backward induction policy \eqref{eq:calibrated_terminal} with $\gamma = 0.975$, per-stage cost $c = 0.001$, and horizon $T = 100$. Figure~\ref{fig:ecmo_bi} shows the resulting stopping boundaries projected onto the posterior mean treatment effect $\hat{\delta}_k$, together with simulated trial paths under the ECMO-like scenario.

Under the ECMO-like scenario ($p_{\text{ECMO}} = 0.80$, $p_{\text{CMT}} = 0.20$), the optimal design stops at a median of 2 patients per arm ($E[N] = 3.0$), with 83\% probability of declaring ECMO superior. Under a moderate effect ($p_{\text{ECMO}} = 0.50$), the median rises to 6 patients per arm. Under the null ($p_{\text{ECMO}} = p_{\text{CMT}} = 0.20$), the type~I error is 0.13---elevated relative to the symmetric-prior simulations of Section~\ref{sec:simulation} because the informative CMT prior concentrates control-arm uncertainty more quickly, increasing the chance that random fluctuations in the ECMO arm produce apparent superiority. This illustrates both the ethical advantage of decision-theoretic designs for large effects and the sensitivity of Bayesian sequential designs to prior asymmetry.

The question of ECMO efficacy was revisited in the EOLIA (Extracorporeal Membrane Oxygenation for Severe Acute Respiratory Distress Syndrome) trial, a separate randomised study of ECMO for adult ARDS (acute respiratory distress syndrome) \citep{goligher2018}. The frequentist p-value was 0.09 (not significant at the conventional 0.05 level), but Bayesian analyses with various priors showed 88--99\% posterior probability that ECMO reduces mortality---a substantially more informative summary than a binary significant/not-significant determination.

\subsection{CALGB 49907: predictive probability monitoring}

The Cancer and Leukemia Group B trial 49907 tested whether capecitabine could replace standard chemotherapy in women with early-stage breast cancer aged 65 or older \citep{muss2009}. The planned sample size ranged from 600 to 1800 patients, with Bayesian predictive probability monitoring for futility and noninferiority at pre-specified interim analyses.

The noninferiority margin corresponded to a 5-year relapse-free survival of 53\% for capecitabine versus 60\% for standard therapy (equivalently, a hazard ratio of~1.24). At each interim analysis, the predictive probability was computed: given current data, what is the probability that future follow-up will yield a conclusive result?

At the first interim analysis after 600 patients, the hazard ratio (standard/capecitabine) was 0.53---equivalently, capecitabine patients had roughly $1/0.53 \approx 1.9$ times the hazard of recurrence. The Bayesian posterior probability that capecitabine failed the noninferiority criterion exceeded 96\%, surpassing the futility threshold of 80\%. Enrolment was discontinued. After an additional year of follow-up, the hazard ratio (capecitabine/standard) was 2.09 (95\% CI: 1.38--3.17; $P < 0.001$), confirming that standard chemotherapy is superior.

The trial achieved its scientific goal with 633 patients instead of the 1800 maximum---a savings of 65\% in sample size and years of accrual time.
\citet{berry2025} describes CALGB~49907 as, to his knowledge, the first fully Bayesian randomised trial published in the \textit{New England Journal of Medicine}.
The design used noninformative priors for treatment effects; the Bayesian contribution was entirely in the decision framework (predictive probabilities for stopping), not in the priors.

\subsection{Platform trials and the Bayesian Time Machine}

Platform trials test multiple treatments within a single protocol, with staggered entry and exit of experimental arms sharing a common control. A central question is whether analyses for a given treatment arm should use only concurrent controls or also non-concurrent controls enrolled earlier in the platform.

\citet{saville2022} proposed the Bayesian Time Machine: a hierarchical model that smooths control-arm response rates across time, allowing analyses to use all control data while adjusting for temporal drift. The model includes time effects as random effects with a smoothing prior that penalises abrupt changes, so that closer time intervals are modelled with similar response rates.

In simulations, the Bayesian Time Machine provides estimates with smaller mean squared error than either concurrent-only or pooled-controls analyses. The smoothing prior acts as a compromise: when temporal drift is absent, it borrows efficiently across all time periods; when drift is present, it limits borrowing to adjacent periods. The I-SPY~2 platform trial for neoadjuvant breast cancer \citep{barker2009} employed this approach across 23 experimental therapies, of which nine graduated to Phase III based on exceeding an 85\% Bayesian predictive probability of confirmatory trial success \citep{berry2025}.

\section{The Regulatory Landscape}\label{sec:regulatory}

The FDA's January 2026 draft guidance on Bayesian methodology \citep{fda2026} specifies concrete requirements for Bayesian submissions:
\begin{enumerate}
\item[(a)] pre-specification of Bayesian success criteria and decision thresholds;
\item[(b)] evaluation of operating characteristics through simulation, including type~I error under the null hypothesis;
\item[(c)] justification of prior distributions, with quantification of their influence (e.g., effective sample size);
\item[(d)] sensitivity analyses examining robustness to prior specification and prior-data conflict;
\item[(e)] documentation sufficient for regulatory reproducibility.
\end{enumerate}

The guidance identifies specific settings where Bayesian methods have supported successful submissions: borrowing from previous trials (e.g., REBYOTA for recurrent \textit{C.\ difficile}), augmenting concurrent controls with external data (e.g., GBM~AGILE for glioblastoma, Precision Promise for pancreatic cancer), paediatric extrapolation (e.g., empagliflozin, linagliptin for paediatric type 2 diabetes), and dose-finding trials in oncology.

The guidance can be read as accommodating two regulatory paths for Bayesian trials. The first requires demonstration of frequentist type~I error control through simulation; the second may permit direct posterior probability thresholds provided that the prior and utility function are pre-specified and agreed upon with the agency, though to the authors' knowledge no approved submission has yet relied solely on this path. Most successful Bayesian submissions to date have followed the first path, using predictive probability monitoring or prior borrowing calibrated to control type~I error. The backward induction framework of this paper fits the second path: the utility function \eqref{eq:utility} and declaration threshold $\gamma$ are pre-specified, and the power frontier (Figure~\ref{fig:power_frontier}) provides exactly the simulation-based operating characteristics the guidance requires under items (a) and~(b). Whether the inflated type~I error inherent in utility-optimal designs (Section~\ref{sec:simulation}) is acceptable remains a regulatory judgement that will depend on the clinical context---rare diseases and paediatric extrapolation, where the guidance explicitly endorses Bayesian methods, are the most natural settings.

A common misconception equates Bayesian methods primarily with informative priors for borrowing. In fact, CALGB~49907 used noninformative priors \citep{muss2009}, and I-SPY~2 used minimally informative treatment-effect priors overwhelmed by accumulating data \citep{berry2025}. In both cases the Bayesian contribution was the decision framework---predictive probabilities for stopping and adaptive enrichment---not the priors. Predictive probabilities, while related to frequentist conditional power, answer a different question: given all observed data, what is the probability of eventual trial success?

\section{Discussion}\label{sec:discussion}

The central theme of this paper is that Bayesian methods can reach strong conclusions with small samples---and with informative priors, even smaller ones---though the sample-size savings come with trade-offs in frequentist operating characteristics that depend on the utility function and cost parameters (Section~\ref{sec:simulation}). The mechanism is the same in every example: the posterior precision equals the sum of prior precision and data precision, so each observation moves the posterior toward a decision boundary, and an informative prior gives a head start. When evidence accumulates in favour of one arm, the trial stops early; when evidence accumulates against it, the trial stops early for futility. Either way, patients are spared. The ECMO trial reached a definitive conclusion with 12 patients. CALGB~49907 stopped at 633 instead of 1800---a 65\% saving. The multi-centre topical cream analysis produced 98.5\% posterior probability of treatment efficacy from centres with as few as five patients per arm by borrowing strength across sites. In I-SPY~2, adaptive enrichment across biomarker-defined subtypes graduated nine of 23 experimental arms to Phase~III, four of which have received marketing approval. In each case, the Bayesian framework converted prior knowledge and accumulating data into rapid decisions that a fixed-sample design could not have reached without enrolling far more patients.

These applied findings rest on three methodological pillars that are consequences of a single principle: probability is the language of uncertainty, and decisions under uncertainty should maximise expected utility. First, priors are not optional nuisance parameters but expressions of genuine knowledge. The fact that treatment effects are small is prior information as real as the trial data; ignoring it via Jeffreys' or uniform priors is not objectivity but a specific modelling choice with known deficiencies (Section~\ref{sec:priors}). The conditional means prior framework of \citet{bedrick1996} provides a principled way to encode this knowledge for regression models, and the P\'olya-Gamma augmentation of \citet{polson2013} removes the historical computational barrier to exact Bayesian inference for the logistic models most commonly used in trial analysis. Second, the posterior updates after every patient, and each update should inform the next decision. Thompson's (1933) allocation rule is the simplest expression of this idea; the backward induction framework of \citet{christen2003} and \citet{carlin1998} provides the optimal rule under the assumed model and loss function, made tractable by the sufficient-statistic reduction.

The exact binary backward induction of Section~\ref{sec:binary} extends this to the two-arm setting most common in practice, while the P\'olya-Gamma bridge (Section~\ref{sec:pgbridge}) connects it to covariate-adjusted logistic regression through a conditional Gaussian structure. Third, clinical trials involve multiple objectives---efficacy, toxicity, information gain, cost---that the utility-based framework of \citet{thall2004} trades off explicitly, producing designs that can improve both patient outcomes and statistical information.

The simulation study (Section~\ref{sec:simulation}) reveals a tension at the heart of Bayesian optimal design. The uncalibrated backward induction rule stops when the expected benefit of further sampling falls below the per-stage cost, achieving expected sample sizes of 14--26 patients per arm but with low power (0.03--0.18), because the small accumulated sample rarely pushes posterior probabilities past a stringent declaration threshold. The calibrated variant, which embeds the declaration threshold directly in the terminal utility, resolves much of this gap: at the same cost ($c = 5 \times 10^{-4}$), power at the largest treatment effect rises from 0.18 to 0.70 while expected sample size roughly doubles from 14 to 24 patients per arm---still well below the 48--100 required by competitors. Reducing the cost further to $c = 10^{-4}$ pushes power to 0.81 at $E[N] = 29$. However, both calibrated designs inflate type~I error to 0.12--0.14 (versus 0.03 for the uncalibrated rule), because the sequential Bayesian policy provides more opportunities for the posterior probability to exceed $\gamma$ by chance under the null. The power frontier (Figure~\ref{fig:power_frontier}) makes the trade-off explicit and tunable: the per-stage cost $c$ controls the operating point on the E[N]-versus-power curve, and informative priors can partially mitigate type~I error inflation (Table~\ref{tab:prior_sensitivity}). In rare-disease or paediatric settings where each patient has high individual value, an operating point with low $E[N]$ may be preferred; in confirmatory trials seeking regulatory approval, predictive probability monitoring and group sequential designs offer tighter type~I error control with power close to the fixed-sample benchmark. The principal contribution of the backward induction framework is therefore not that it outperforms these alternatives on frequentist criteria, but that it delivers the optimal policy for a stated utility---exact in the uncalibrated Beta-Binomial recursion, approximate in the calibrated variant where a Normal approximation to $\Pr(p_1 > p_0 \mid \text{data})$ gates the terminal reward---and makes the resulting trade-offs---between sample size, power, and error rate---explicit and tunable.

Backward induction is best understood as a design option on the power--sample-size frontier, suited to patient-sparing settings such as rare diseases, paediatrics, and early adaptive phases, rather than a replacement for confirmatory group sequential or fixed-sample designs. In regulatory practice, $c$ and $\gamma$ should be calibrated by simulation to meet pre-specified type~I error and power targets before the trial begins, as required by the FDA guidance \citep{fda2026}. The case studies of Section~\ref{sec:cases} carry the practical argument for Bayesian methods; the simulation reveals the theoretical limits of decision-theoretic optimality when judged by frequentist operating characteristics.

Several limitations should be noted. Our analysis draws on published trial summaries rather than patient-level data, so the case studies illustrate rather than independently replicate the original findings. The Normal model backward induction assumes known observation variance; unknown variance adds a second sufficient statistic and a correspondingly larger grid. The binary backward induction is exact on the integer lattice but scales as $O(T^3)$, limiting practical horizons to a few hundred patients per arm; for larger trials, approximate methods such as predictive probability monitoring are more tractable. The P\'olya-Gamma bridge to logistic regression (Section~\ref{sec:pgbridge}) relies on a Laplace approximation at the posterior mode of the augmented variables. Table~\ref{tab:pg_validation} shows that this approximation is accurate across the sample sizes tested (mean absolute error below 0.01 for $n \geq 10$ per arm), though accuracy at very small samples ($n < 10$) or with complete separation should be verified against exact calculations. The conjugate and logistic models that make P\'olya-Gamma augmentation and conditional means priors tractable may not capture the complexities of time-to-event, longitudinal, or competing-risk endpoints that arise in modern platform trials. Finally, we do not address the practical challenges of prior elicitation from clinical teams---a step that remains the principal barrier to routine adoption of informative priors.

The 2026 FDA guidance \citep{fda2026} now provides a concrete regulatory pathway, and the accumulating record of Bayesian adaptive trials in oncology, rare diseases, and platform settings suggests that the remaining barriers are practical---prior elicitation, software infrastructure, and institutional familiarity---rather than methodological.

\subsection*{Data and code availability}
No patient-level data were used in this paper. All figures are reproducible from the R scripts provided as supplementary material.

\subsection*{Competing interests}
The authors declare no competing interests.

\subsection*{Author contributions}
A.S.\ contributed clinical trial expertise and case study material. V.S.\ developed the mathematical framework, computational examples, and wrote the manuscript. N.P.\ provided the decision-theoretic perspective and editorial guidance.

\bibliography{ref}
\end{document}